\documentclass[aps,prd,nofootinbib,amsmath,amssymb,superscriptaddress,showkeys,showpacs,twocolumn,10pt]{revtex4}

\usepackage{graphicx}
\usepackage{dcolumn}
\usepackage{bm}
\usepackage{amssymb}
\usepackage{latexsym}
\usepackage{booktabs}
\usepackage{amsmath}
\usepackage{multirow}
\usepackage[colorlinks=true, linkcolor=red, citecolor=blue]{hyperref}

\def\be{\begin{equation}}
\def\ee{\end{equation}}
\def\ba{\begin{eqnarray}}
\def\ea{\end{eqnarray}}

\begin{document}

\title{Constraints on active and sterile neutrinos in an interacting dark energy cosmology}

\author{Lu Feng}
\affiliation{College of Physical Science and Technology, Shenyang Normal University, Shenyang
110034, China}
\author{Dong-Ze He}
\affiliation{Department of Physics, College of Sciences, Northeastern University, Shenyang 110819, China}
\author{Hai-Li Li}
\affiliation{Department of Physics, College of Sciences, Northeastern University, Shenyang 110819, China}
\author{Jing-Fei Zhang}
\affiliation{Department of Physics, College of Sciences, Northeastern University, Shenyang 110819, China}
\author{Xin Zhang\footnote{Corresponding author}}
\email{zhangxin@mail.neu.edu.cn}
\affiliation{Department of Physics, College of Sciences, Northeastern University, Shenyang 110819, China}
\affiliation{Ministry of Education's Key Laboratory of Data Analytics and Optimization for Smart Industry, Northeastern University, Shenyang 110819, China}
\affiliation{Center for High Energy Physics, Peking University, Beijing 100080, China}

\begin{abstract}
We investigate the impacts of dark energy on constraining massive (active/sterile) neutrinos in interacting dark energy (IDE) models by using the current observations. We employ two typical IDE models, the interacting $w$ cold dark matter (I$w$CDM) model and the interacting holographic dark energy (IHDE) model, to make an analysis. To avoid large-scale instability, we use the parameterized post-Friedmann approach to calculate the cosmological perturbations in the IDE models. The cosmological observational data used in this work include the Planck cosmic microwave background (CMB) anisotropies data, the baryon acoustic oscillation data, the type Ia supernovae data, the direct measurement of the Hubble constant, the weak lensing data, the redshift-space distortion data, and the CMB lensing data. We find that the dark energy properties could influence the constraint limits of active neutrino mass and sterile neutrino parameters in the IDE models. We also find that the dark energy properties could influence the constraints on the coupling strength parameter $\beta$, and a positive coupling constant, $\beta>0$, can be detected at the $2.5\sigma$ statistical significance for the IHDE+$\nu_s$ model by using the all-data combination. In addition, we also discuss the ``Hubble tension'' issue in these scenarios. We find that the $H_0$ tension can be effectively relieved by considering massive sterile neutrinos, and in particular in the IHDE+$\nu_s$ model the $H_0$ tension can be reduced to be at the $1.28\sigma$ level.

\end{abstract}


\maketitle

\section{Introduction}
\label{sec1}
The solar and atmospheric neutrino oscillation experiments have revealed that neutrinos have masses and there is a significant mixing between different neutrino species (for reviews, see, e.g., Refs.~\cite{Lesgourgues:2006nd,Xing:2019vks}).
In addition, anomalies of short-baseline and reactor neutrino experiments suggest that there could be a sterile neutrino with eV-scale mass~\cite{Abazajian:2012ys,Hannestad:2012ky,Conrad:2013mka}.
To directly measure the absolute neutrino masses for active neutrinos and search for sterile neutrinos, different experiments are needed.
In fact, the cosmological observations have provided the tightest limits on massive (active/sterile) neutrinos.

Using observations of the cosmic microwave background (CMB) from Planck satellite~\cite{Aghanim:2015xee}, the baryon acoustic oscillations (BAO)~\cite{Beutler:2011hx,Ross:2014qpa,Cuesta:2015mqa}, the type Ia supernovae (SN)~\cite{Betoule:2014frx}, the Hubble constant ($H_0$)~\cite{Riess:2016jrr}, and the CMB lensing~\cite{Ade:2015zua}, a constraint of $\sum m_\nu<0.197$~eV (2$\sigma$) has been achieved in the case of active neutrinos~\cite{Zhang:2015uhk}, whereas in the sterile neutrino case we have $N_{\rm eff}<3.572$ and $m_{\nu,{\rm sterile}}^{\rm eff}<0.268$~eV  (2$\sigma$) \cite{Feng:2017mfs} in the $\Lambda$CDM model. For the dynamical dark energy models, the $w$CDM model gives $\sum m_\nu<0.304$~eV (2$\sigma$) for the active neutrinos, and $N_{\rm eff}<3.501$ and $m_{\nu,{\rm sterile}}^{\rm eff}<0.531$~eV (2$\sigma$) for the sterile neutrinos. In addition, the holographic dark energy (HDE) model gives $\sum m_\nu<0.113$~eV (2$\sigma$) for the active neutrinos \cite{Zhang:2015uhk}, and $N_{\rm eff}<3.670$ and $m_{\nu,{\rm sterile}}^{\rm eff}<0.199$~eV (2$\sigma$) for the sterile neutrinos \cite{Feng:2017mfs}, and thus the HDE model leads to the extremely stringent upper limit on the total mass of active neutrinos and the effective mass of sterile neutrinos.
For other relevant studies of massive (active/sterile) neutrinos by using cosmological data, see, e.g., Refs.~\cite{GonzalezGarcia:2007ib,Li:2012vn,Wang:2012uf,Wang:2012vh,Mirizzi:2013gnd,Drewes:2013gca,Wyman:2013lza,Hamann:2013iba,Battye:2013xqa,Giusarma:2014zza,Zhang:2014dxk,
Dvorkin:2014lea,Archidiacono:2014apa,Zhang:2014nta,Leistedt:2014sia,Costanzi:2014tna,Zhang:2014lfa,Zhu:2014qma,Li:2015poa,Zhang:2015rha,DellOro:2015kys,
Huang:2015wrx,Capozzi:2016rtj,Giusarma:2016phn,DiValentino:2016hlg,Lu:2016hsd,Wang:2016tsz,Zhao:2016ecj,Vagnozzi:2017ovm,Guo:2017hea,Li:2017iur,
Capozzi:2017ipn,Feng:2017nss,Zhao:2017urm,Zhao:2017jma,Feng:2017usu,Guo:2018gyo,Guo:2018ans,Zhao:2018fjj,Feng:2019mym,Jacques:2013xr,Lesgourgues:2014zoa,Geng:2015haa,Qian:2015waa,
Patterson:2015xja,Gariazzo:2015rra,Xu:2016ddc,Yang:2017amu,Abazajian:2017tcc,Wang:2017htc,Wang:2018lun,Zhang:2019ipd}.

In addition, when a direct interaction between dark energy (DE) and cold dark matter (CDM) is considered in current cosmology, the constraints become $\sum m_\nu<0.10$~eV (2$\sigma$) in the $Q=\beta H\rho_{\rm de}$ model and $\sum m_\nu<0.20$~eV (2$\sigma$) in the $Q=\beta H\rho_{\rm c}$ model for the case of active neutrinos \cite{Guo:2017hea}, as well as $N_{\rm eff}<3.641$ and $m_{\nu,{\rm sterile}}^{\rm eff}<0.312$~eV (2$\sigma$) in the $Q=\beta H\rho_{\rm de}$ model, and $N_{\rm eff}<3.498$ and $m_{\nu,{\rm sterile}}^{\rm eff}<0.875$~eV (2$\sigma$) in the $Q=\beta H\rho_{\rm c}$ model for the case of sterile neutrinos \cite{Feng:2017usu}. These results imply that the interaction has an important impacts on neutrino parameters.
Moreover, in Refs.~\cite{Guo:2017hea,Feng:2017usu} only the study of interacting vacuum energy model (i.e., the I$\Lambda$CDM model) is done, but the dark energy properties are not considered. Actually, dark energy properties would also lead to important impact  on the constraints on the neutrino parameters \cite{Zhang:2015uhk,Feng:2017mfs,Feng:2019mym}. Thus, in this paper, we will revisit the constraints on the neutrinos parameters in interacting dark energy (IDE) models (for IDE, see e.g. Refs.~\cite{Comelli:2003cv,Cai:2004dk,Zhang:2004gc,Zhang:2005rg,Zimdahl:2005bk,Wang:2006qw,Guo:2007zk,Zhang:2007uh,He:2008tn,Li:2009zs,Zhang:2009qa,He:2010im,Chen:2011rz,
Li:2011ga,Clemson:2011an,Fu:2011ab,Zhang:2012uu,Salvatelli:2013wra,Xu:2013jma,Zhang:2013zyn,Li:2013bya,Yang:2014gza,yang:2014vza,Geng:2015ara,Duniya:2015nva,
Yin:2015pqa,Murgia:2016ccp,Wang:2016lxa,Pourtsidou:2016ico,Costa:2016tpb,Feng:2016djj,Xia:2016vnp,vandeBruck:2016hpz,Sola:2017jbl,Yang:2018euj,Guo:2017deu,Li:2018ydj}).

The IDE model considers DE directly interacting with CDM by exchanging energy and momentum. To investigate the IDE model, one starts from the continuity equations for DE and CDM by assuming an energy transfer between them, in the background cosmology,

\begin{align}
&\rho'_{\rm de} = -3\mathcal{H}(1+w)\rho_{\rm de}+ aQ_{\rm de},\label{eq1}\\
&\rho'_{\rm c} = -3\mathcal{H}\rho_{\rm c}+aQ_{\rm c},~~~~~~Q_{\rm de}=-Q_{\rm c}=Q,\label{eq2}
\end{align}
where $\rho_{\rm de}$ and $\rho_{\rm c}$ are energy densities of DE and CDM, respectively, $\mathcal{H}=a'/a$ ($a$ is the scale factor of the universe) is the conformal Hubble parameter, a prime is the derivative with respect to the conformal time $\eta$, $w$ is the equation of state (EoS) parameter of DE, and $Q$ is the energy transfer rate. In this work, we consider the phenomenological form of $Q=\beta H_0 \rho_{\rm c}$, where $\beta$ is the coupling constant. From Eqs. (\ref{eq1}) and~(\ref{eq2}), we see that $\beta>0$ means that the energy transfer is from CDM to DE, $\beta<0$ means that the energy transfer is from DE to CDM, and $\beta=0$ means that there is no interaction between DE and CDM.

In this paper, we focus on investigating the impacts of interacting dynamical dark energy on constraining the active neutrino total mass and sterile neutrino parameters.
Following the studies of Ref.~\cite{Feng:2019mym} (as well as Refs.~\cite{Zhang:2015uhk,Feng:2017mfs} without the consideration of interaction in the DE models), we consider two typical dynamical dark energies in the IDE models.
The first one we choose is the interacting $w$ cold dark matter (I$w$CDM) model, which is the interacting version of the $w$CDM cosmology. In the $w$CDM model, the dark energy has a constant EoS parameter $w$.
The other one we choose is the interacting holographic dark energy (IHDE) model, which is the interacting version of the HDE cosmology. In the HDE model, the dark energy density is of the form $\rho_{\rm de}=3c^2 M^2_{\rm pl} R_{\rm{EH}}^{-2}$, where $c$ is a dimensionless parameter that determines the evolution of dark energy, $M_{\rm pl}$ is the reduced Planck mass, and $R_{\rm{EH}}$ is the event horizon of the universe. The evolution of EoS of HDE is given by $w(a)=-1/3-(2/3c)\sqrt{\Omega_{\rm de}(a)}$. For more details of the HDE model, see, e.g., Refs~\cite{Zhang:2015rha,Wang:2016och,Bamba:2012cp,Zhang:2005hs,Chang:2005ph,Zhang:2006av,Zhang:2006qu,Zhang:2007sh,Zhang:2007uh,Zhang:2007es,Zhang:2007an,Ma:2007av,Li:2009bn,Zhang:2009xj,Cui:2010dr,Wang:2012uf}.
Combining Eqs. (\ref{eq1}), (\ref{eq2}) and the EoS of the $w$CDM and HDE cosmologies, the background evolutions of the I$w$CDM and IHDE models can be obtained.
In this paper, we constrain the IDE (I$w$CDM and IHDE) models with massive (active/sterile) neutrinos by using the current cosmological observations, and we report the constraint results and make a detailed analysis for them.

This paper is organized as follows. In Sec.~\ref{sec2}, we present the analysis method and the observational data used in this paper. In Sec. \ref{sec3}, we report the constraint results and discuss these results in detail. Conclusion is given in Sec. \ref{sec4}.

\section{Method and data}
\label{sec2}
In this paper, we place constraints on the massive (active/sterile) neutrinos in IDE cosmology with the current observational data.

The free parameter vector for the IDE model is $\{\omega_b,~\omega_c,~100\theta_{\rm MC},~\tau,~n_s,~\ln (10^{10}A_s),~w~({\rm or}~c),~\beta\},$
where $\omega_b=\Omega_bh^2$ is the physical baryon density, $\omega_c=\Omega_ch^2$ is the physical cold dark matter density, $\theta_{\rm MC}$ is the ratio (multiplied by 100) of the sound horizon to angular diameter distance at decoupling, $\tau$ is the reionization optical depth, $n_s$ is the scalar spectral index, $A_s$ is the amplitude of primordial scalar perturbation power spectrum, $w$ is the EoS parameter of dark energy in the I$w$CDM model, $c$ is the parameter determining the evolution of dark energy in the IHDE model, and $\beta$ is the coupling parameter characterizing the interaction strength between cold dark matter and dark energy.

When we further consider massive neutrinos in cosmology, the total neutrino mass $\sum m_\nu$ should be added for the case of active neutrinos and the parameters $m_{\nu,{\rm sterile}}^{\rm eff}$ (the effective sterile neutrino mass) and $N_{\rm eff}$ (the effective number of relativistic species) should be added for the case of sterile neutrinos.
When the active neutrino mass is considered in the I$\Lambda$CDM, I$w$CDM, and IHDE models, these cases are called the I$\Lambda$CDM+$\nu_a$, I$w$CDM+$\nu_a$, and IHDE+$\nu_a$ models, respectively. Similarly, when the sterile neutrino species is considered in the I$\Lambda$CDM, I$w$CDM, and IHDE models, these cases are called the I$\Lambda$CDM+$\nu_s$, I$w$CDM+$\nu_s$, and IHDE+$\nu_s$ models, respectively.
Thus, the I$\Lambda$CDM+$\nu_a$ model has eight independent parameters and I$\Lambda$CDM+$\nu_s$ model has nine independent parameters. Likewise, the I$w$CDM+$\nu_a$ and IHDE+$\nu_a$ models have nine base parameters, and the I$w$CDM+$\nu_s$ and IHDE+$\nu_s$ models have ten base parameters.

The IDE cosmology actually was troubled by the perturbation instability~\cite{Zhao:2005vj,Valiviita:2008iv,He:2008si} and this problem had greatly frustrated the advance of the IDE cosmology.
Therefore, how to treat the perturbation instability becomes a crucial issue in the current study of IDE cosmology. For the conventional way, the calculation of the dark energy has induced several instabilities, which arises from the incorrect way of calculating the dark energy pressure perturbation, as pointed out in Refs.~\cite{Zhao:2005vj,Valiviita:2008iv,He:2008si}. In current situation, we can handle the instability problem of the IDE cosmology based on the parametrized post-Friedmann (PPF) approach~\cite{Li:2014eha,Li:2014cee,Li:2015vla,Feng:2017usu} (for the original version of PPF, see Refs.~\cite{Hu:2008zd,Fang:2008sn}). With the help of PPF, we can explore the whole parameter space.
To infer the posterior probability distributions of the parameters, our constraints are based on the latest version of the Markov-Chain Monte Carlo package {\tt CosmoMC} ~\cite{Lewis:2002ah}.

The observational data sets we use in this paper are comprised of CMB, BAO, SN, $H_0$, RSD,  WL, and CMB lensing.
\begin{itemize}
\item  The CMB data: We use the CMB anisotropies data from the {\it Planck} 2015 release including TT, TE, EE, and lowP data~\cite{Aghanim:2015xee}.
\item  The BAO data: We employ the baryon acoustic oscillation data from the 6dFGS measurement at $z_{\rm eff}=0.106$~\cite{Beutler:2011hx}, the SDSS-MGS measurement at $z_{\rm eff}=0.15$~\cite{Ross:2014qpa}, and the LOWZ and CMASS samples from BOSS DR12 at $z_{\rm eff}=0.32$ and $z_{\rm eff}=0.57$~\cite{Cuesta:2015mqa}, respectively.
\item  The SN data: For the type Ia supernova observation (SN), we use the Joint Light-curve Analysis (JLA) sample~\cite{Betoule:2014frx}.
\item  The $H_0$ data: We use the Hubble constant direct measurement $H_0=73.24{\pm1.74}~{\rm km}~{\rm s}^{-1}~{\rm Mpc}^{-1}$~\cite{Riess:2016jrr}.
\item  The RSD data: We employ two redshift space distortion (RSD) measurements from the LOWZ and CMASS samples from BOSS DR12 at $z_{\rm eff}=0.32$ and $z_{\rm eff}=0.57$~\cite{Gil-Marin:2016wya}, respectively.
\item  The WL data: We use the cosmic shear measurement of weak lensing (WL) from the CFHTLenS survey~\cite{Heymans:2013fya}.
\item  The lensing data: We also use the CMB lensing power spectrum from the {\it Planck} lensing measurement~\cite{Ade:2015zua}.
\end{itemize}

In this paper, we will use these observational data sets to place constraints on the IDE models with massive (active/sterile) neutrinos. We will compare the I$w$CDM and IHDE models with the I$\Lambda$CDM model under the utilization of uniform data combinations. In our analysis, we use two data combinations, i.e., CMB+BAO+SN+$H_0$ and  CMB+BAO+SN+$H_0$+WL+RSD+lensing. For convenience, we use the abbreviations ``CMB+BSH" and ``CMB+BSH+LSS"  to denote these two combinations, respectively. In what follows, we will report and discuss the fitting results of the IDE models with massive neutrinos in light of these two data combinations.
\section{Results and discussion}\label{sec3}

\begin{table*}\small
\setlength\tabcolsep{1.5pt}
\renewcommand{\arraystretch}{1.5}
\caption{\label{tabactive}Fitting results for the I$\Lambda$CDM+$\nu_a$, I$w$CDM+$\nu_a$, and IHDE+$\nu_a$ models from the data combinations CMB+BSH and CMB+BSH+LSS. Best fit values with $\pm1\sigma$ errors are presented, but for the parameter $\sum m_\nu$, the $2\sigma$ upper limits are given.}
\centering
\begin{tabular}{cccccccccccc}
\hline \multicolumn{1}{c}{Model} &&\multicolumn{2}{c}{I$\Lambda$CDM+$\nu_a$}&&\multicolumn{2}{c}{I$w$CDM+$\nu_a$}&&\multicolumn{2}{c}{IHDE+$\nu_a$}&\\
           \cline{1-1}\cline{3-4}\cline{6-7}\cline{9-10}

 Data  && CMB+BSH &CMB+BSH+LSS&& CMB+BSH &CMB+BSH+LSS &&CMB+BSH &CMB+BSH+LSS\\
\hline

$\Omega_{\rm m}$
                      &&$0.277\pm0.016$
                      &$0.295\pm0.014$
                      &&$0.307^{+0.022}_{-0.026}$
                      &$0.282^{+0.018}_{-0.019}$
                      &&$0.249^{+0.017}_{-0.021}$
                      &$0.255^{+0.016}_{-0.018}$\\

$\sigma_8$
                      &&$0.862\pm0.023$
                      &$0.816\pm0.015$
                      &&$0.850\pm0.018$
                      &$0.816^{+0.014}_{-0.013}$
                      &&$0.843\pm0.018$
                      &$0.828\pm0.012$\\

$H_0\,[{\rm km}/{\rm s}/{\rm Mpc}]$
                      &&$69.39\pm0.84$
                      &$68.44\pm0.70$
                      &&$69.55^{+0.93}_{-0.94}$
                      &$68.82^{+0.84}_{-0.85}$
                      &&$69.89^{+0.95}_{-0.94}$
                      &$69.14^{+0.85}_{-0.84}$\\

$\beta$
                      &&$0.084^{+0.051}_{-0.057}$
                      &$0.032^{+0.042}_{-0.051}$
                      &&$-0.055\pm0.094$
                      &$0.084^{+0.079}_{-0.089}$
                      &&$0.195^{+0.093}_{-0.095}$
                      &$0.191^{+0.079}_{-0.081}$\\

$w$
                      &&...
                      &...
                      &&$-1.114^{+0.098}_{-0.081}$
                      &$-0.994^{+0.068}_{-0.059}$
                      &&...
                      &...\\

$c$
                      &&...
                      &...
                      &&...
                      &...
                      &&$0.766^{+0.081}_{-0.099}$
                      &$0.784^{+0.075}_{-0.084}$\\

$\sum m_\nu\,[\rm eV]$
                      &&$<0.214$
                      &$<0.269$
                      &&$<0.161$
                      &$<0.319$
                      &&$<0.125$
                      &$<0.151$\\
\hline
\end{tabular}

\end{table*}

\begin{table*}\small
\setlength\tabcolsep{1.5pt}
\renewcommand{\arraystretch}{1.5}
\caption{\label{tabsterile}Fitting results for the I$\Lambda$CDM+$\nu_s$, I$w$CDM+$\nu_s$, and IHDE+$\nu_s$ models from the data combinations CMB+BSH and CMB+BSH+LSS. Best fit values with $\pm1\sigma$ errors are presented, but for the parameters $N_{\rm eff}$ and $m_{\nu,{\rm{sterile}}}^{\rm{eff}}$, the $2\sigma$ upper limits are given.}
\centering
\begin{tabular}{cccccccccccc}
\hline \multicolumn{1}{c}{Model} &&\multicolumn{2}{c}{I$\Lambda$CDM+$\nu_s$}&&\multicolumn{2}{c}{I$w$CDM+$\nu_s$}&&\multicolumn{2}{c}{IHDE+$\nu_s$}&\\
           \cline{1-1}\cline{3-4}\cline{6-7}\cline{9-10}

 Data  && CMB+BSH &CMB+BSH+LSS&& CMB+BSH &CMB+BSH+LSS &&CMB+BSH &CMB+BSH+LSS\\
\hline

$\Omega_{\rm m}$
                      &&$0.279^{+0.016}_{-0.019}$
                      &$0.284^{+0.015}_{-0.017}$
                      &&$0.310^{+0.025}_{-0.029}$
                      &$0.278^{+0.018}_{-0.020}$
                      &&$0.250^{+0.019}_{-0.023}$
                      &$0.251^{+0.016}_{-0.018}$\\

$\sigma_8$
                      &&$0.845^{+0.031}_{-0.026}$
                      &$0.790^{+0.024}_{-0.021}$
                      &&$0.839^{+0.031}_{-0.021}$
                      &$0.786^{+0.025}_{-0.022}$
                      &&$0.828^{+0.031}_{-0.021}$
                      &$0.799^{+0.027}_{-0.022}$\\

$H_0\,[{\rm km}/{\rm s}/{\rm Mpc}]$
                      &&$70.13^{+0.96}_{-1.05}$
                      &$69.68^{+0.89}_{-0.98}$
                      &&$70.30^{+1.10}_{-1.20}$
                      &$70.00\pm1.00$
                      &&$70.60\pm1.10$
                      &$70.05\pm0.97$\\

$\beta$
                      &&$0.087\pm0.059$
                      &$0.087\pm0.060$
                      &&$-0.060^{+0.100}_{-0.110}$
                      &$0.105^{+0.084}_{-0.085}$
                      &&$0.200\pm0.110$
                      &$0.223^{+0.091}_{-0.089}$\\

$w$
                      &&...
                      &...
                      &&$-1.109^{+0.100}_{-0.084}$
                      &$-1.007^{+0.066}_{-0.058}$
                      &&...
                      &...\\

$c$
                      &&...
                      &...
                      &&...
                      &...
                      &&$0.771^{+0.084}_{-0.108}$
                      &$0.771^{+0.074}_{-0.085}$\\

$m_{\nu,{\rm{sterile}}}^{\rm{eff}}\,[\rm eV]$
                                             &&$<0.647$
                                             &$0.460^{+0.210}_{-0.350}$
                                             &&$<0.569$
                                             &$0.430^{+0.190}_{-0.280}$
                                             &&$<0.662$
                                             &$<0.927$\\

$N_{\rm{eff}}$
                     &&$<3.516$
                     &$3.259^{+0.082}_{-0.172}$
                     &&$<3.591$
                     &$3.253^{+0.079}_{-0.164}$
                     &&$<3.574$
                     &$3.186^{+0.034}_{-0.138}$\\
\hline

\end{tabular}

\end{table*}

\begin{figure*}[!htp]
\begin{center}
\includegraphics[width=12.0cm]{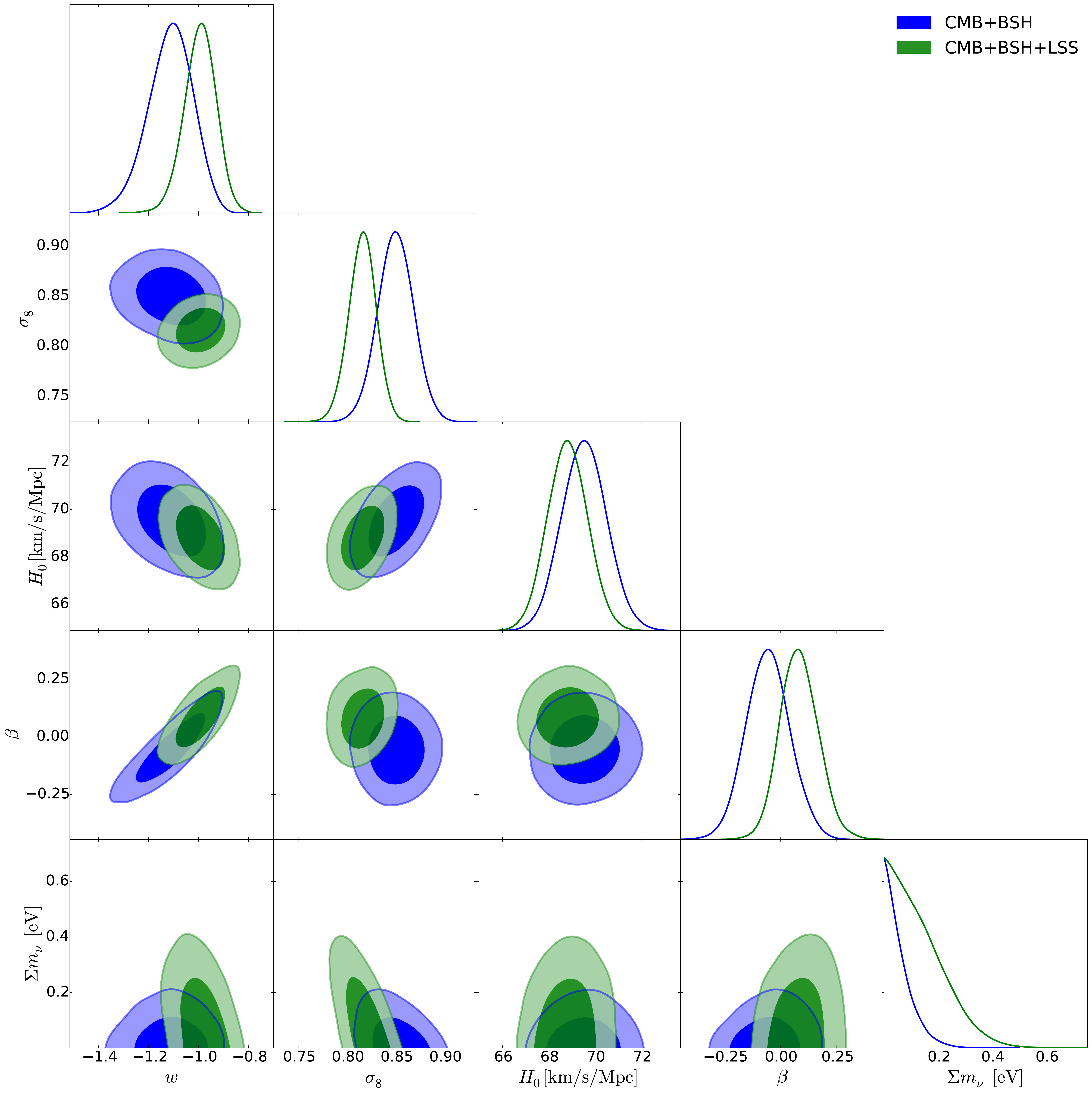}
\end{center}
\caption{\label{vaiw}(Color online) The one-dimensional posterior distributions and two-dimensional marginalized contours (1$\sigma$ and 2$\sigma$) for parameters $w$, $\sigma_8$, $H_0$, $\beta$, and $\sum m_\nu$ of the I$w$CDM+$\nu_a$ model by using the CMB+BSH ({\it blue}) and the CMB+BSH+LSS ({\it green}) data combinations.}
\end{figure*}

\begin{figure*}[!htp]
\begin{center}
\includegraphics[width=12.0cm]{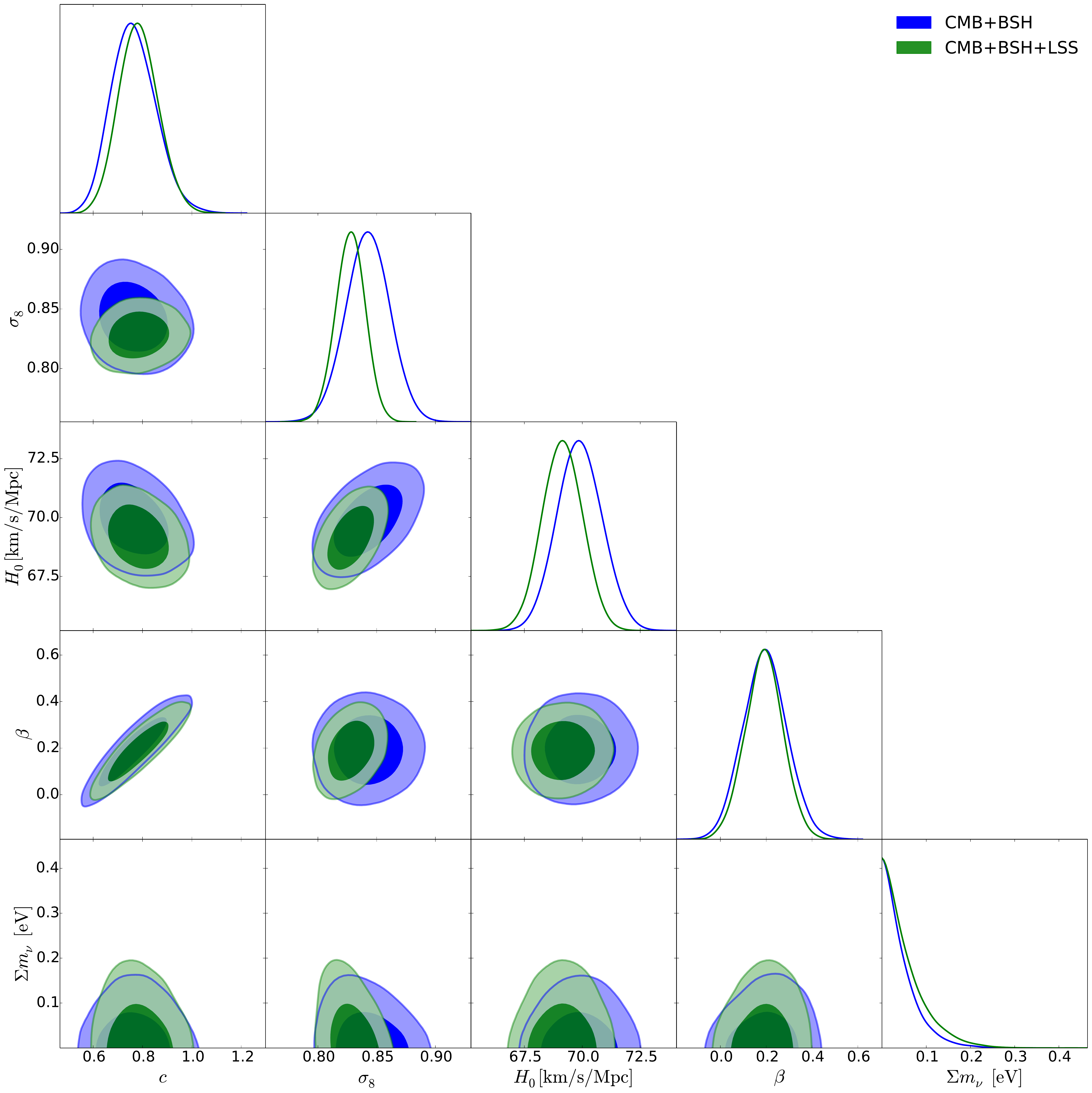}
\end{center}
\caption{\label{vaih} (Color online) The one-dimensional posterior distributions and two-dimensional marginalized contours (1$\sigma$ and 2$\sigma$) for parameters $c$, $\sigma_8$, $H_0$, $\beta$, and $\sum m_\nu$ of the IHDE+$\nu_a$ model by using the CMB+BSH ({\it blue}) and the CMB+BSH+LSS ({\it green}) data combinations.}
\end{figure*}

\begin{figure*}[!htp]
\begin{center}
\includegraphics[width=8.0cm]{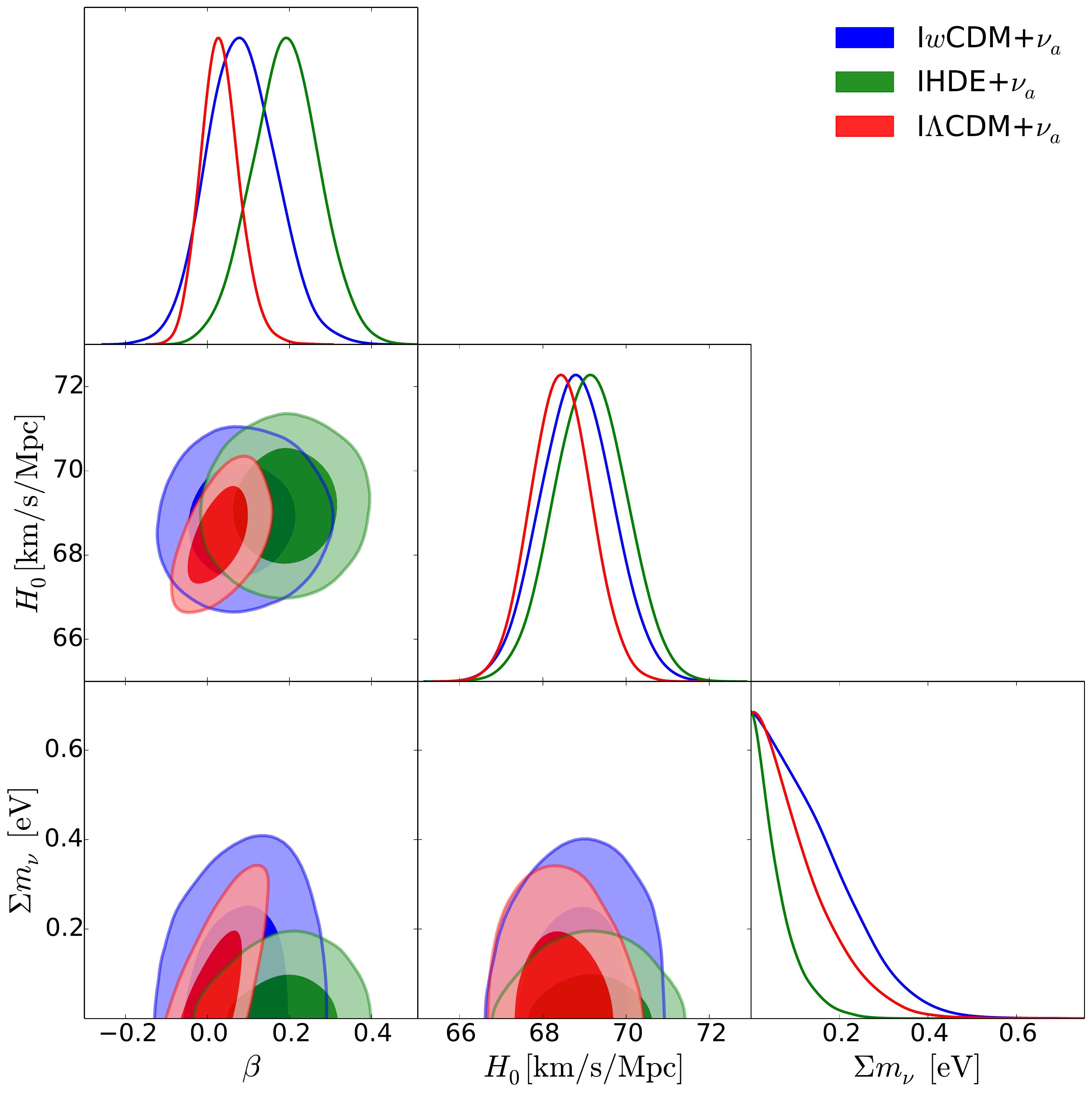}
\end{center}
\caption{\label{vaiwh} (Color online) The one-dimensional posterior distributions and two-dimensional marginalized contours (1$\sigma$ and 2$\sigma$) for parameters $\beta$, $H_0$, and $\sum m_\nu$ of the I$\Lambda$CDM+$\nu_a$, I$w$CDM+$\nu_a$, and IHDE+$\nu_a$ models by using the CMB+BSH+LSS data combination.}
\end{figure*}

\begin{figure*}[!htp]
\begin{center}
\includegraphics[width=12.0cm]{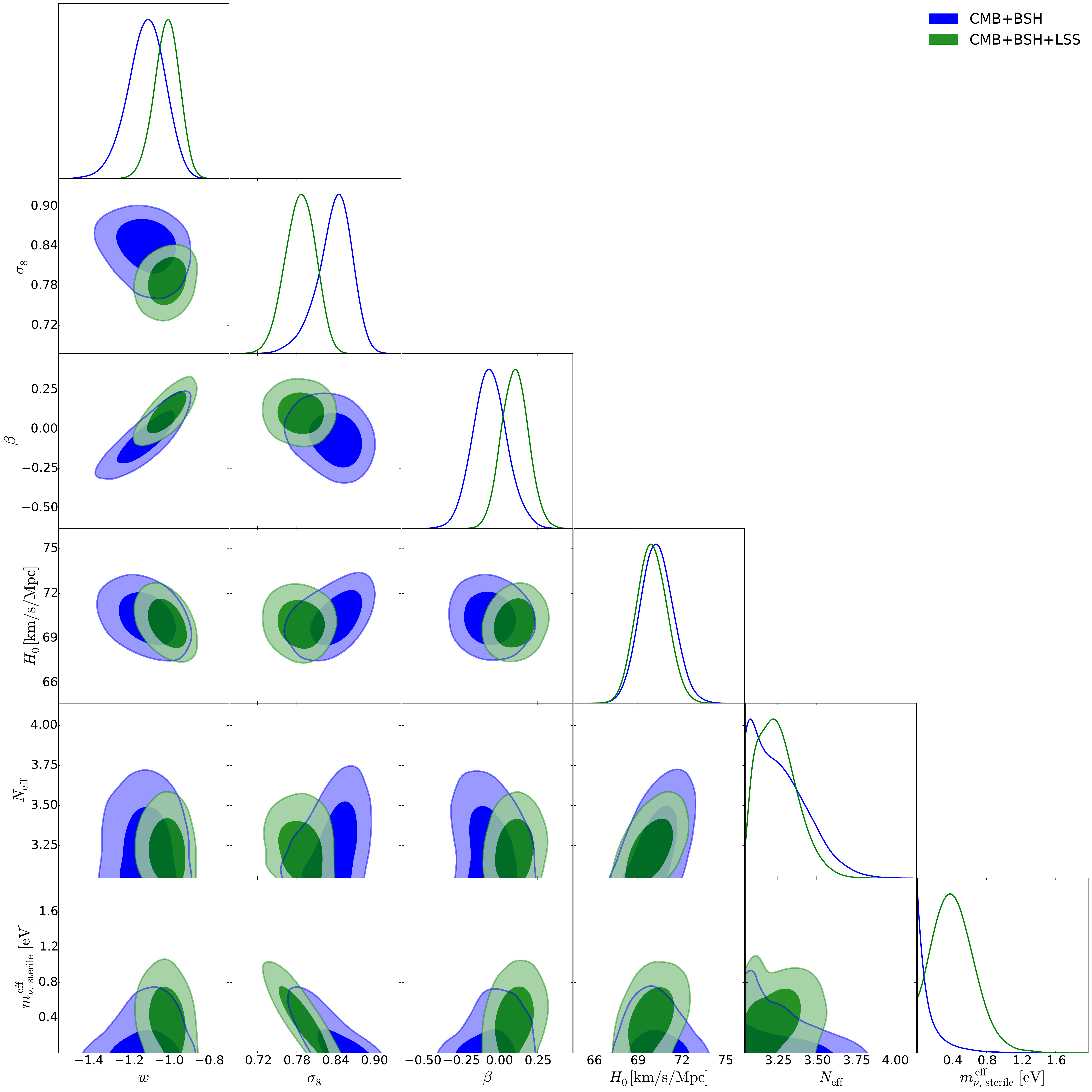}
\end{center}
\caption{\label{vsiw} (Color online) The one-dimensional posterior distributions and two-dimensional marginalized contours (1$\sigma$ and 2$\sigma$) for parameters $w$, $\sigma_8$, $\beta$, $H_0$, $N_{\rm eff}$, and $m_{\nu,{\rm sterile}}^{\rm eff}$ of the I$w$CDM+$\nu_s$ model by using the CMB+BSH ({\it blue}) and the CMB+BSH+LSS ({\it green}) data combinations.}
\end{figure*}

\begin{figure*}[!htp]
\begin{center}
\includegraphics[width=12.0cm]{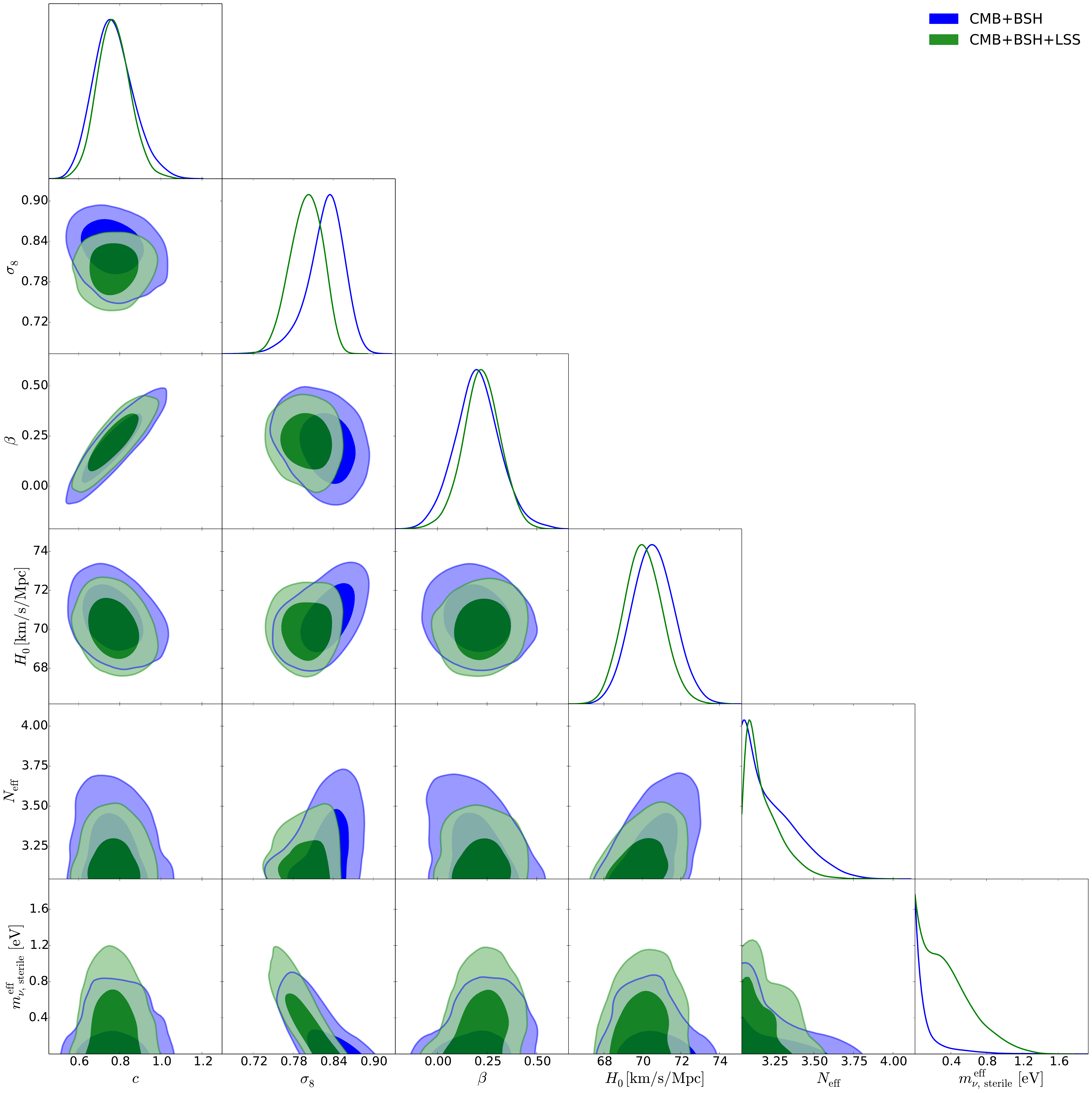}
\end{center}
\caption{\label{vsih} (Color online) The one-dimensional posterior distributions and two-dimensional marginalized contours (1$\sigma$ and 2$\sigma$) for parameters $c$, $\sigma_8$, $\beta$, $H_0$, $N_{\rm eff}$, and $m_{\nu,{\rm sterile}}^{\rm eff}$ of the IHDE+$\nu_s$ model by using the CMB+BSH ({\it blue}) and the CMB+BSH+LSS ({\it green}) data combinations.}
\end{figure*}

\begin{figure*}[!htp]
\begin{center}
\includegraphics[width=8.0cm]{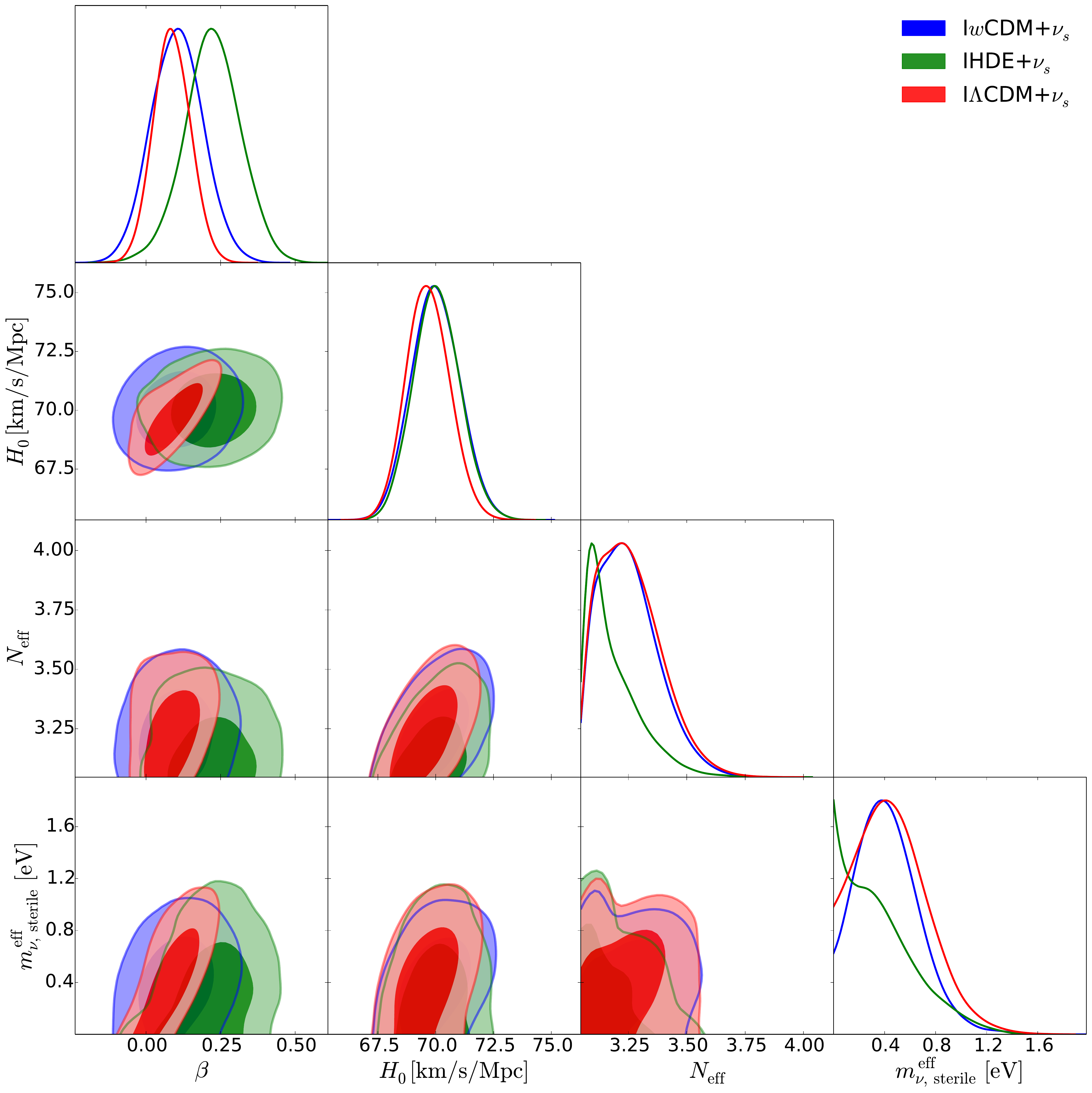}
\end{center}
\caption{\label{vsiwh} (Color online) The one-dimensional posterior distributions and two-dimensional marginalized contours (1$\sigma$ and 2$\sigma$) for parameters $\beta$, $H_0$, $N_{\rm eff}$, and $m_{\nu,{\rm sterile}}^{\rm eff}$ of the I$\Lambda$CDM+$\nu_s$, I$w$CDM+$\nu_s$, and IHDE+$\nu_s$ models by using the CMB+BSH+LSS data combination.}
\end{figure*}

\begin{figure*}[!htp]
\begin{center}
\includegraphics[width=5.5cm]{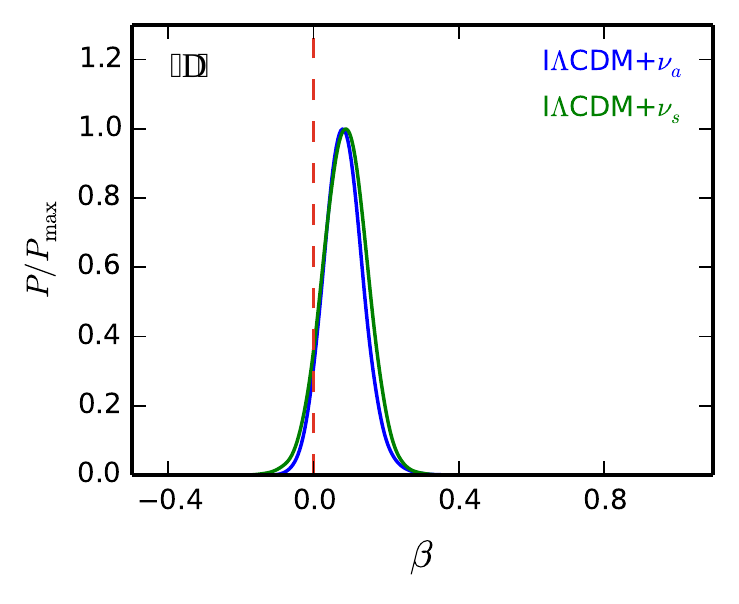}
\includegraphics[width=5.5cm]{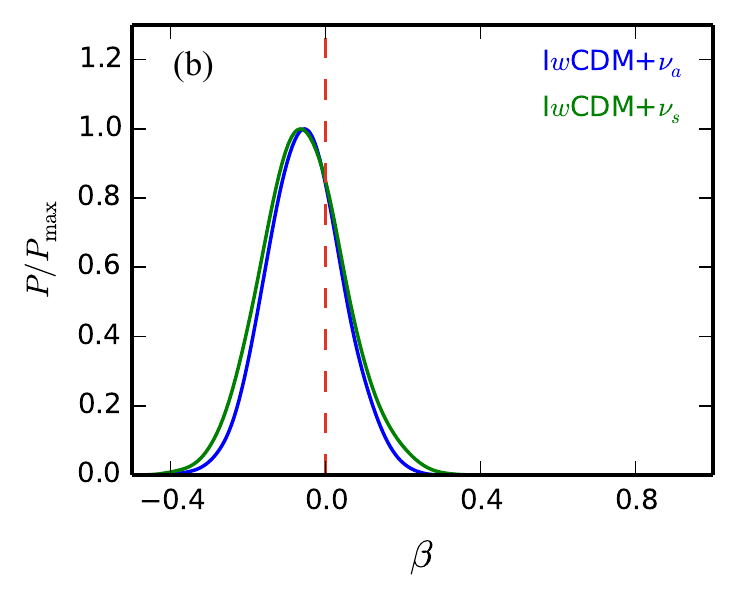}
\includegraphics[width=5.5cm]{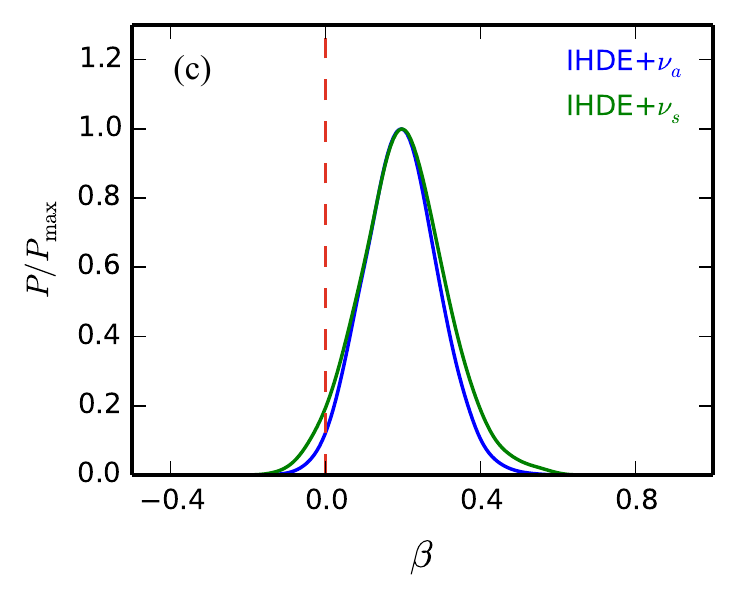}
\includegraphics[width=5.5cm]{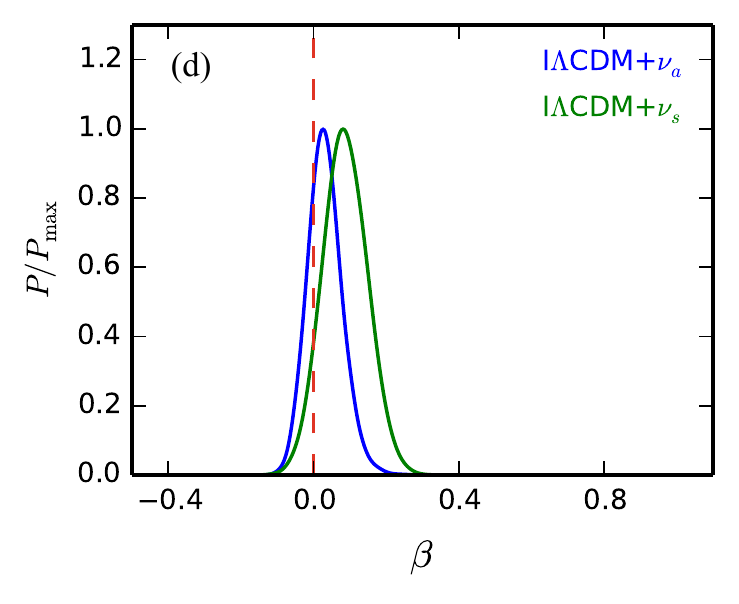}
\includegraphics[width=5.5cm]{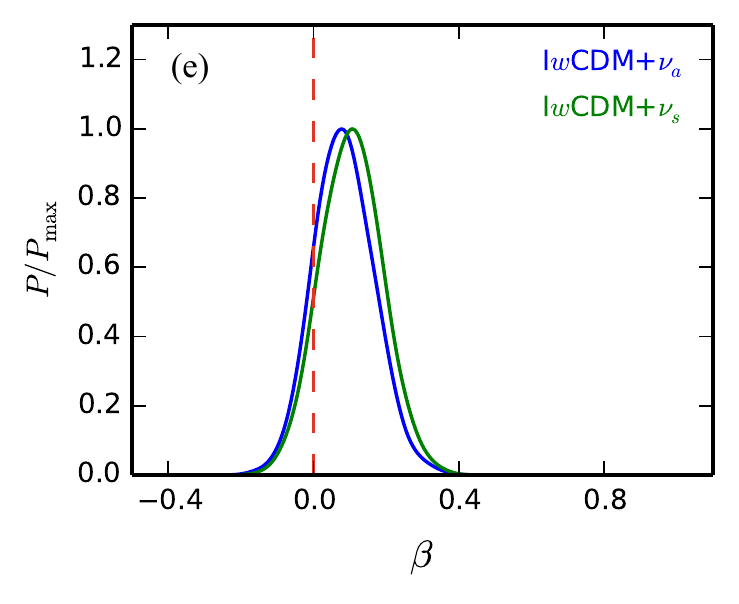}
\includegraphics[width=5.5cm]{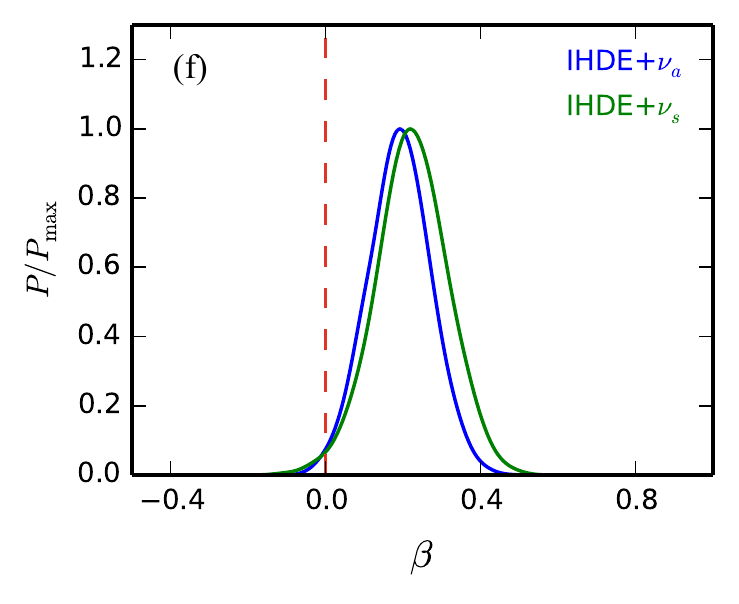}
\end{center}
\caption{\label{likeb} (Color online) The one-dimensional posterior distributions for the coupling parameter $\beta$ for the IDE models by using CMB+BSH (a)-(c) and CMB+BSH+LSS (d)-(f).}
\end{figure*}

\begin{figure*}[!htp]
\begin{center}
\includegraphics[width=7.0cm]{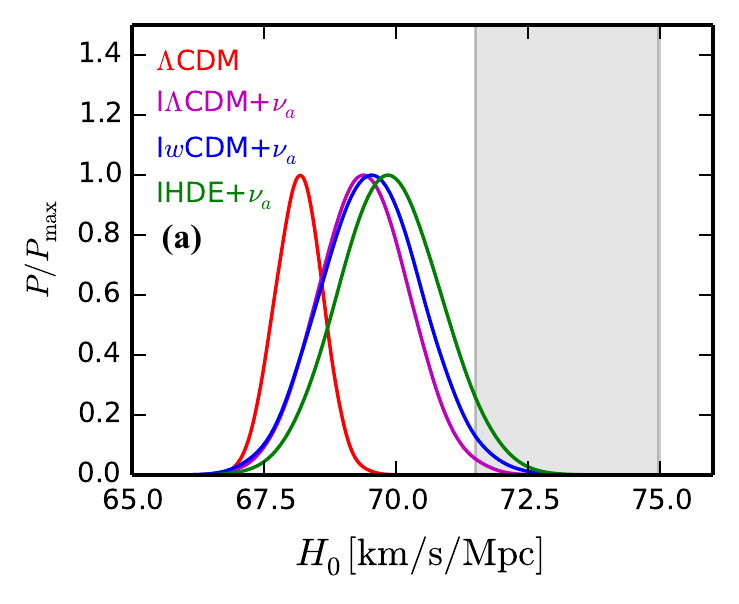}
\includegraphics[width=7.0cm]{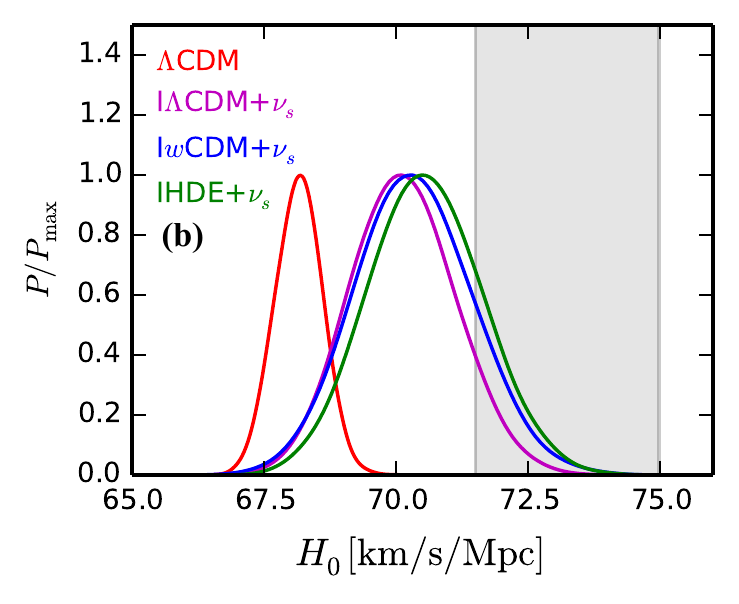}
\includegraphics[width=7.0cm]{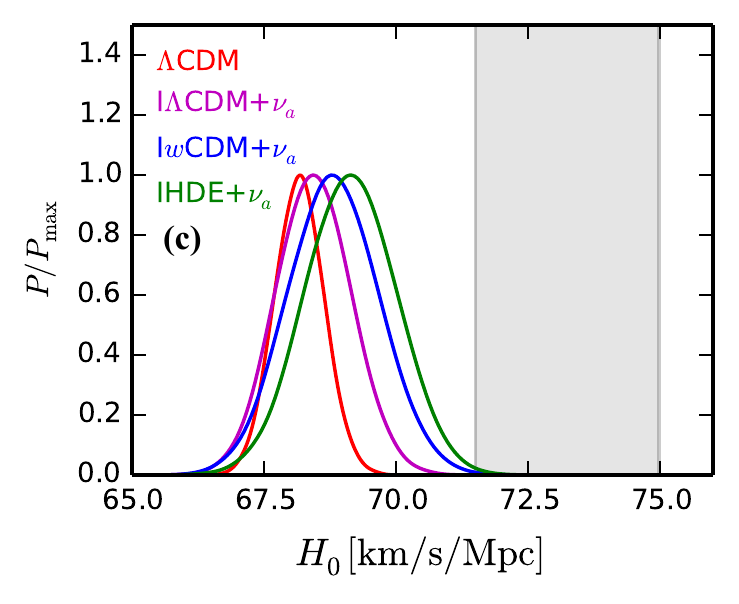}
\includegraphics[width=7.0cm]{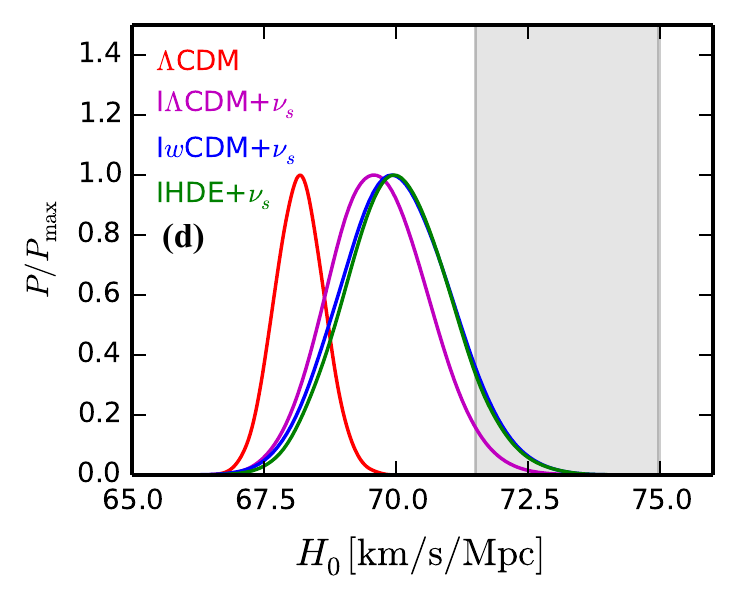}
\end{center}
\caption{\label{likeh01} (Color online) The one-dimensional posterior distributions for the parameter $H_0$ for the $\Lambda$CDM and IDE models by using CMB+BSH (a) and (b) and CMB+BSH+LSS (c) and (d). The result of the local measurement of Hubble constant ($H_0=73.24\pm1.74~{\rm km}~{\rm s}^{-1}~{\rm Mpc}^{-1}$) is shown by the grey band.}
\end{figure*}

\begin{figure*}[!htp]
\begin{center}
\includegraphics[width=5.5cm]{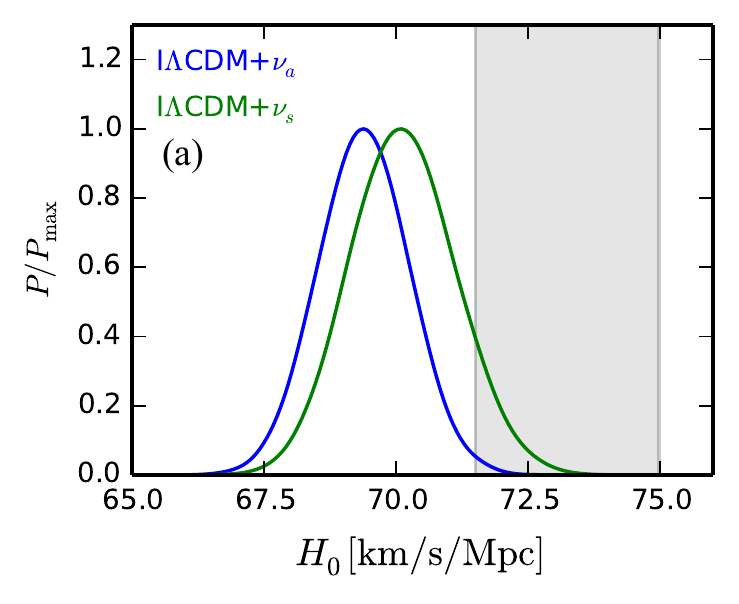}
\includegraphics[width=5.5cm]{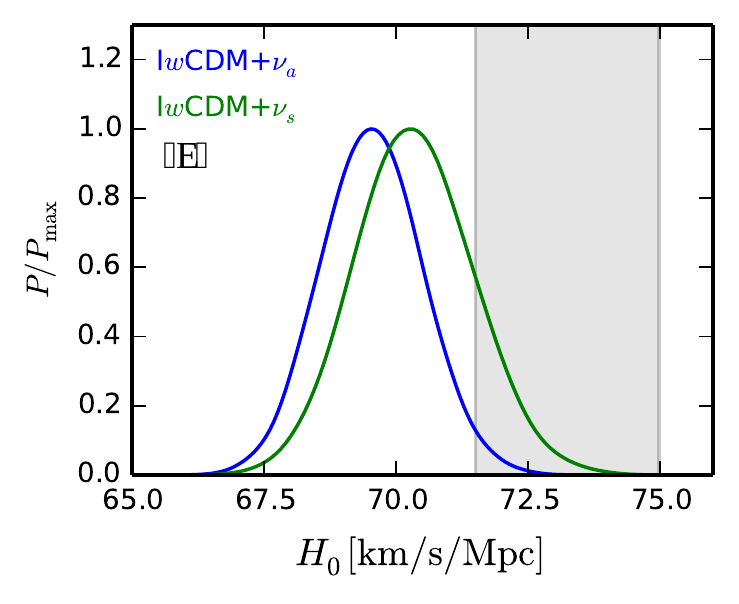}
\includegraphics[width=5.5cm]{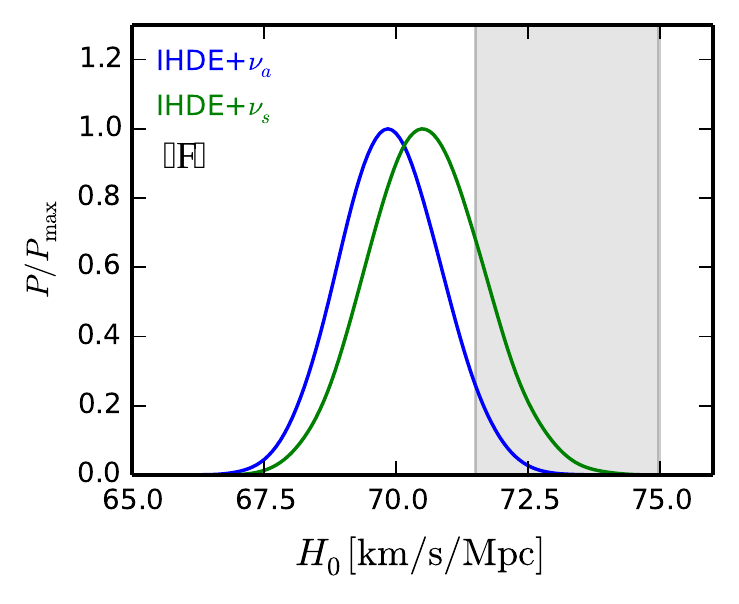}
\includegraphics[width=5.5cm]{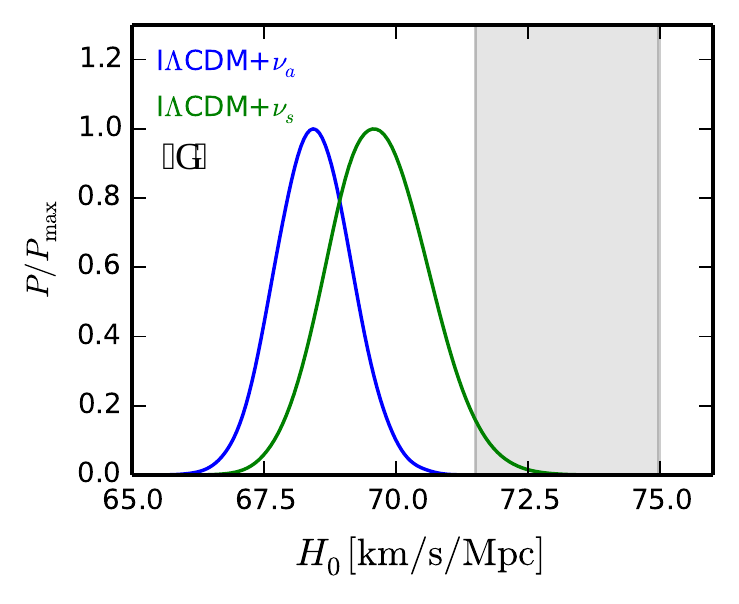}
\includegraphics[width=5.5cm]{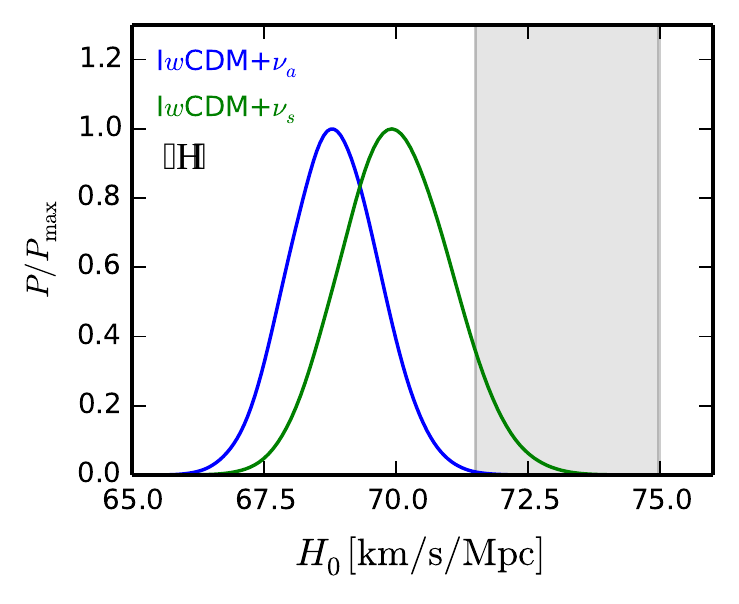}
\includegraphics[width=5.5cm]{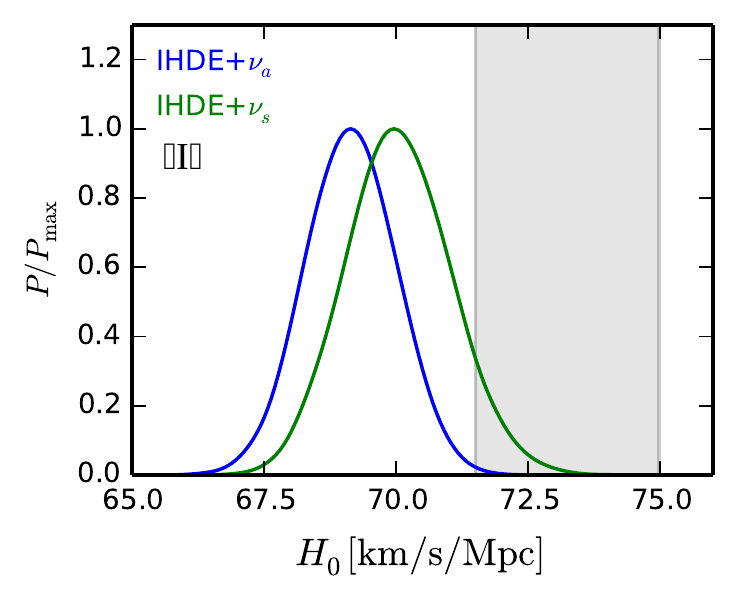}
\end{center}
\caption{\label{likeh02} (Color online) The one-dimensional posterior distributions for the parameter $H_0$ for the IDE models by using CMB+BSH (a)-(c) and CMB+BSH+LSS (d)-(f). The result of the local measurement of Hubble constant ($H_0=73.24\pm1.74~{\rm km}~{\rm s}^{-1}~{\rm Mpc}^{-1}$) is shown by the grey band.}
\end{figure*}
In this section, we shall report the fitting results of the overall considered interacting dynamical dark energy models and discuss the implications of the results. We use the CMB+BSH and CMB+BSH+LSS data combinations to constrain these models. The fitting results are displayed in Tables~\ref{tabactive}--\ref{tabsterile} and Figs.~\ref{vaiw}--\ref{likeh02}.

\subsection{Active neutrino mass}
In this subsection, we present the constraint results of the I$w$CDM+$\nu_a$ and the IHDE+$\nu_a$ models by using the CMB+BSH and CMB+BSH+LSS data combinations and analyze how the dark energy properties affect the constraints on the active neutrino mass in the IDE models. As we take the I$\Lambda$CDM+$\nu_a$ as a reference model in this work, the fitting results of this model from the same data combinations are also given. The main results can be seen in Table~\ref{tabactive} and Figs.~\ref{vaiw}--\ref{vaiwh}.

For the I$w$CDM+$\nu_a$ model, we obtain $\sum m_\nu<0.161$~eV from the data combination CMB+BSH, and $\sum m_\nu<0.319$~eV from CMB+BSH+LSS.
We find that, compared with the I$\Lambda$CDM+$\nu_a$ model, the upper limits on the parameter $\sum m_\nu$ become slightly tighter from CMB+BSH (this result has been discussed in detail in Ref.~\cite{Feng:2019mym}), while for the case of CMB+BSH+LSS, the constraints on $\sum m_\nu$ would be somewhat looser.
For the I$w$CDM+$\nu_a$ model, the one-dimensional posterior distributions and two-dimensional marginalized contours (1$\sigma$ and 2$\sigma$) for the parameters $w$, $\sigma_8$, $H_0$, $\beta$, and $\sum m_\nu$ with the data combinations of CMB+BSH and CMB+BSH+LSS are shown in Fig.~\ref{vaiw}.
One can clearly see that, $\sum m_\nu$ is in slightly positive-correlation with $w$ for the data combination of CMB+BSH, while $\sum m_\nu$ and $w$ are in anti-correlation for the case of CMB+BSH+LSS. Evidently, when adding the LSS (WL+RSD+lensing) data, the correlation between $\sum m_\nu$ and $w$ could be significantly changed, leading to a different constraints on the active neutrino mass $\sum m_\nu$.

For the IHDE+$\nu_a$ model, we obtain $\sum m_\nu<0.125$~eV from the data combination CMB+BSH, and $\sum m_\nu<0.151$~eV from CMB+BSH+LSS. Clearly, the IHDE+$\nu_a$ model provides the most stringent upper limits on $\sum m_\nu$ among the three models, i.e., I$\Lambda$CDM+$\nu_a$, I$w$CDM+$\nu_a$, and IHDE+$\nu_a$, for the data combinations of CMB+BSH and CMB+BSH+LSS (see also Ref.~\cite{Feng:2019mym}), which is also accordant with the conclusion in the previous studies without the consideration of interaction between dark sectors~\cite{Zhang:2015uhk,Feng:2017mfs,Wang:2016tsz}.
The one- and two-dimensional marginalized posterior distributions of the parameters $c$, $\sigma_8$, $H_0$, $\beta$, and $\sum m_\nu$ for the IHDE+$\nu_a$ model using CMB+BSH and CMB+BSH+LSS are shown in Fig.~\ref{vaih}.

In addition, we can notice that in these three models, the addition of the LSS data would lead to a larger neutrino mass limit (see Table~\ref{tabactive}). This is because the LSS observation favors a lower matter perturbations ($\sigma_8$), and neutrino mass $\sum m_\nu$ is anti-correlated with $\sigma_8$ due to the free-streaming effect of active neutrinos (see Figs.~\ref{vaiw} and~\ref{vaih}), thus a larger neutrino mass limit could be derived. This result is also accordant with the conclusion in the previous study~\cite{Guo:2017hea}.

To directly show how the dark energy properties affect the constraint on the active neutrino mass, in Fig.~\ref{vaiwh} we further make a comparison for the overall three IDE models (i.e., I$\Lambda$CDM+$\nu_a$, I$w$CDM+$\nu_a$, and IHDE+$\nu_a$) with the data combination of CMB+BSH+LSS. From this figure, we find that in the I$w$CDM+$\nu_a$ model, the limits on $\sum m_\nu$ become much looser, while in the IHDE+$\nu_a$ model the limits become much more stringent. Thus, we can conclude that the dark energy properties could influence the constraint results of the active neutrino mass $\sum m_\nu$.

\subsection{Sterile neutrino parameters}
In this subsection, we shall present the constraint results of sterile neutrino parameters in the three IDE models. The main results are given in Table~\ref{tabsterile} and Figs.~\ref{vsiw}--\ref{vsiwh}.

For the I$w$CDM+$\nu_s$ model, we obtain $m_{\nu,{\rm{sterile}}}^{\rm{eff}}<0.569$~eV and $N_{\rm{eff}}<3.591$ from the data combination CMB+BSH, and $m_{\nu,{\rm{sterile}}}^{\rm{eff}}=0.430^{+0.190}_{-0.280}$~eV and $N_{\rm{eff}}=3.253^{+0.079}_{-0.164}$ from CMB+BSH+LSS.
We find that, in the I$w$CDM+$\nu_s$ model, the fitting results of $m_{\nu,{\rm{sterile}}}^{\rm{eff}}$ and $N_{\rm{eff}}$ are basically consistent with those of the I$\Lambda$CDM+$\nu_s$ model regardless of using any of these two data combinations (see in Table~\ref{tabsterile}). In addition,  when the LSS (WL+RSD+lensing) data are considered in the cosmological fit, the addition of the LSS data could significantly improve the constraints on the sterile neutrino parameters ($m_{\nu,{\rm{sterile}}}^{\rm{eff}}$ and $N_{\rm{eff}}$). Our results prefer a non-zero mass at the 1.314$\sigma$ level and $\Delta N_{\rm{eff}}>0$ at the 1.238$\sigma$ level for the I$\Lambda$CDM+$\nu_s$ model and prefer a non-zero mass at the 1.536$\sigma$ level and $\Delta N_{\rm{eff}}>0$ at the 1.262$\sigma$ level for the I$w$CDM+$\nu_s$ model.
This indicates that the LSS data could play an important role in constraining the sterile neutrino parameters. These results are also in accordance with the conclusions in the previous studies~\cite{Feng:2017nss,Zhao:2017urm}. The one-dimensional posterior distributions and two-dimensional marginalized contours (1$\sigma$ and 2$\sigma$) for the parameters $w$, $\sigma_8$, $\beta$, $H_0$, $N_{\rm eff}$, and $m_{\nu,{\rm sterile}}^{\rm eff}$ of the I$w$CDM+$\nu_s$ model using both the two data combinations are shown in Fig.~\ref{vsiw}.

For the IHDE+$\nu_s$ model, we obtain $m_{\nu,{\rm{sterile}}}^{\rm{eff}}<0.662$~eV and $N_{\rm{eff}}<3.574$ from CMB+BSH. We find that, in the IHDE+$\nu_s$ model, the fitting results of $m_{\nu,{\rm{sterile}}}^{\rm{eff}}$ and $N_{\rm{eff}}$ are also basically consistent with those of the I$\Lambda$CDM+$\nu_s$ model (same as the I$w$CDM+$\nu_s$ model). This is because the data combination CMB+BSH is not sensitive to the sterile neutrino parameters and thus can not give a tight constraint on the parameters of sterile neutrino. These results are in accordance with the conclusions in the previous studies for the models without dark sectors' interaction~\cite{Feng:2017mfs}.  Moreover, the CMB+BSH+LSS constraint gives  $m_{\nu,{\rm{sterile}}}^{\rm{eff}}<0.927$~eV and $N_{\rm{eff}}=3.186^{+0.034}_{-0.138}$. Evidently, the addition of the LSS (WL+RSD+lensing) data would lead to a larger upper limit of the effective sterile neutrino mass $m_{\nu,{\rm{sterile}}}^{\rm{eff}}$. This is in that a lower $\sigma_8$ is favored by the current LSS observations, and $m_{\nu,{\rm{sterile}}}^{\rm{eff}}$ is anti-correlated with $\sigma_8$. Thus, a larger effective sterile neutrino mass upper limit could be derived (see Fig.~\ref{vsih}). This result is accordant with the case of massive active neutrino (in this study) and also accordant with the conclusion in the previous studies~\cite{Feng:2017usu}.

To directly show how the dark energy properties affect the constraint on sterile neutrino parameters, we plot the comparison of these three models, i.e., I$\Lambda$CDM+$\nu_s$, I$w$CDM+$\nu_s$, and IHDE+$\nu_s$, from CMB+BSH+LSS combination, in Fig.~\ref{vsiwh}.
We find that, in the I$w$CDM+$\nu_s$ model, the one-dimensional posterior distribution curves and two-dimensional marginalized contours in $m_{\nu,{\rm{sterile}}}^{\rm{eff}}$--$N_{\rm{eff}}$ plane are almost in coincidence with the I$\Lambda$CDM+$\nu_s$ model. However, in the IHDE+$\nu_s$ model, the constraint results for these two parameters are evidently different. Thus, we can conclude that the dark energy properties could also influence the constraints on the sterile neutrino parameters.

\subsection{Coupling parameter}
In this subsection, we discuss the fitting results of the coupling constant $\beta$. The main results are given in Tables~\ref{tabactive}--\ref{tabsterile} and Figs.~\ref{vaiw}--\ref{likeb}.

In Table~\ref{tabactive}, we show the constraint results of the I$\Lambda$CDM+$\nu_a$, I$w$CDM+$\nu_a$, and IHDE+$\nu_a$ models from the CMB+BSH and CMB+BSH+LSS data combinations.
By using CMB+BSH data, we obtain $\beta=0.084^{+0.051}_{-0.057}$ for the I$\Lambda$CDM+$\nu_a$ model, $\beta=-0.055\pm0.094$ for the I$w$CDM+$\nu_a$ model, and $\beta=0.195^{+0.093}_{-0.095}$ for the IHDE+$\nu_a$ model.
By using CMB+BSH+LSS data, we obtain $\beta=0.032^{+0.042}_{-0.051}$ for the I$\Lambda$CDM+$\nu_a$ model, $\beta=0.084^{+0.079}_{-0.089}$ for the I$w$CDM+$\nu_a$ model, and $\beta=0.191^{+0.079}_{-0.081}$ for the IHDE+$\nu_a$ model.
We find that, compared with the I$\Lambda$CDM+$\nu_a$ model, the fitting results of $\beta$ in the I$w$CDM+$\nu_a$ and IHDE+$\nu_a$ models are quite different.

For the I$w$CDM+$\nu_a$ model, CMB+BSH data combination favors a negative coupling constant, which means that the dark energy decays into dark matter. However, CMB+BSH+LSS data combination favors a positive coupling constant, which implies that dark matter decays into dark energy.
For the IHDE+$\nu_a$ model, both of the two data combinations consistently favor a positive coupling constant, indicating that the case of dark matter decaying into dark energy is more supported in this scenario.
Moreover, for both of the two data combinations, $\beta=0$ is favored within the $1\sigma$ significance range for the I$w$CDM+$\nu_a$ model and $\beta>0$ is favored at more than $2\sigma$ significance level for the IHDE+$\nu_a$ model (see also Figs.~\ref{vaiw}--\ref{vaiwh}).

In Table~\ref{tabsterile}, we show the constraint results of the I$\Lambda$CDM+$\nu_s$, I$w$CDM+$\nu_s$, and IHDE+$\nu_s$ models from the CMB+BSH and CMB+BSH+LSS data combinations.
By using the CMB+BSH combination, we obtain $\beta=0.087\pm0.059$ for the I$\Lambda$CDM+$\nu_s$ model, $\beta=-0.060^{+0.100}_{-0.110}$ for the I$w$CDM+$\nu_s$ model, and $\beta=0.200\pm0.110$ for the IHDE+$\nu_s$ model.
By using the CMB+BSH+LSS combination, we obtain $\beta=0.087\pm0.060$ for the I$\Lambda$CDM+$\nu_s$ model, $\beta=0.105^{+0.084}_{-0.085}$ for the I$w$CDM+$\nu_s$ model, and $\beta=0.223^{+0.091}_{-0.089}$ for the IHDE+$\nu_s$ model.
Obviously, we find that compared with the I$\Lambda$CDM+$\nu_s$ model, the fitting results of $\beta$ in the I$w$CDM+$\nu_s$ and IHDE+$\nu_s$ models are quite different, which is consistent with the conclusions in the studies of the active neutrinos, manifesting that the dark energy properties could play an important role in changing the fitting results of the coupling constant $\beta$. To show the effect of the dark energy properties on the constraints of the coupling parameter $\beta$, the one- and two-dimensional posterior distribution contours of $\beta$ are shown in Figs.~\ref{vsiw}--\ref{vsiwh}.
It is of great interest to find that $\beta>0$ is favored at more than $2\sigma$ (about $2.507\sigma$) level for the IHDE+$\nu_s$ model by using the CMB+BSH+LSS data combination (see also Figs.~\ref{vsih}--\ref{vsiwh}).

In addition, with the purpose of directly showing the impacts of massive (active/sterile) neutrinos on the constraints of the coupling parameter $\beta$, we further perform an analysis for all the IDE models (with active neutrinos or with sterile neutrinos) considered in this paper, to make a comparison. The one-dimensional posterior distributions of $\beta$ for the six models (I$\Lambda$CDM+$\nu_a$, I$\Lambda$CDM+$\nu_s$, I$w$CDM+$\nu_a$, I$w$CDM+$\nu_s$, IHDE+$\nu_a$, and IHDE+$\nu_s$) using the CMB+BSH and CMB+BSH+LSS data combinations are shown in Fig.~\ref{likeb}.
Obviously, we find that, for both of the two data combinations,
the consideration of massive (active/sterile) neutrinos could scarcely influence the constraints on coupling constant $\beta$, which is accordant with the conclusions in the previous studies on the I$\Lambda$CDM model~\cite{Feng:2017usu}.

\subsection{The $H_0$ tension}
In this subsection, we shall discuss the issue of the $H_0$ tension between the {Planck} observation and the Hubble constant direct measurement. The main results are given in Tables~\ref{tabactive}--\ref{tabsterile} and Figs.~\ref{vaiw}--\ref{vsiwh} and~\ref{likeh01}--\ref{likeh02}.

We first constrain the base $\Lambda$CDM model using the CMB+BSH and CMB+BSH+LSS data combinations and obtain $H_0= 68.61^{+0.44}_{-0.45}~{\rm km}~{\rm s}^{-1}~{\rm Mpc}^{-1}$ (CMB+BSH) and $H_0= 68.17^{+0.45}_{-0.46}~{\rm km}~{\rm s}^{-1}~{\rm Mpc}^{-1}$ (CMB+BSH+LSS), which are $2.56 \sigma$ (CMB+BSH) and $2.82 \sigma$ (CMB+BSH+LSS) lower than the distance-ladder measurement of the Hubble constant, $H_0=73.24\pm1.74~{\rm km}~{\rm s}^{-1}~{\rm Mpc}^{-1}$, respectively. In order to obtain a highter $H_0$, we further consider the interacting dark energy cosmology with the addition of massive (active/sterile) neutrinos.

In Tables~\ref{tabactive} and \ref{tabsterile}, we show the constraint results of the I$\Lambda$CDM+$\nu_a$, I$w$CDM+$\nu_a$, IHDE+$\nu_a$, I$\Lambda$CDM+$\nu_s$, I$w$CDM+$\nu_s$, and IHDE+$\nu_s$ models from the CMB+BSH and CMB+BSH+LSS data combinations.
By using the CMB+BSH data combination, we obtain $H_0=69.39\pm0.84~{\rm km}~{\rm s}^{-1}~{\rm Mpc}^{-1}$ for the I$\Lambda$CDM+$\nu_a$ model, $H_0=69.55^{+0.93}_{-0.94}~{\rm km}~{\rm s}^{-1}~{\rm Mpc}^{-1}$ for the I$w$CDM+$\nu_a$ model, $H_0=69.89^{+0.95}_{-0.94}~{\rm km}~{\rm s}^{-1}~{\rm Mpc}^{-1}$ for the IHDE+$\nu_a$ model, $H_0=70.13^{+0.96}_{-1.05}~{\rm km}~{\rm s}^{-1}~{\rm Mpc}^{-1}$ for the I$\Lambda$CDM+$\nu_s$ model, $H_0=70.30^{+1.10}_{-1.20}~{\rm km}~{\rm s}^{-1}~{\rm Mpc}^{-1}$ for the I$w$CDM+$\nu_s$ model, and $H_0=70.60\pm1.10~{\rm km}~{\rm s}^{-1}~{\rm Mpc}^{-1}$ for the IHDE+$\nu_s$ model,
which indicate that the tensions with the local value of the Hubble constant, $H_0=73.24\pm1.74~{\rm km}~{\rm s}^{-1}~{\rm Mpc}^{-1}$, would be reduced to be at the levels of $1.99\sigma$, $1.87\sigma$, $1.69\sigma$, $1.56\sigma$, $1.43\sigma$, and $1.28\sigma$, respectively.
By using the CMB+BSH+LSS data combination, we obtain $H_0=68.44\pm0.70~{\rm km}~{\rm s}^{-1}~{\rm Mpc}^{-1}$ for the I$\Lambda$CDM+$\nu_a$ model, $H_0=68.82^{+0.84}_{-0.85}~{\rm km}~{\rm s}^{-1}~{\rm Mpc}^{-1}$ for the I$w$CDM+$\nu_a$ model, $H_0=69.14^{+0.85}_{-0.84}~{\rm km}~{\rm s}^{-1}~{\rm Mpc}^{-1}$ for the IHDE+$\nu_a$ model, $H_0=69.68^{+0.89}_{-0.98}~{\rm km}~{\rm s}^{-1}~{\rm Mpc}^{-1}$ for the I$\Lambda$CDM+$\nu_s$ model, $H_0=70.00\pm1.00~{\rm km}~{\rm s}^{-1}~{\rm Mpc}^{-1}$ for the I$w$CDM+$\nu_s$ model, and $H_0=70.05\pm0.97~{\rm km}~{\rm s}^{-1}~{\rm Mpc}^{-1}$ for the IHDE+$\nu_s$ model.
In this case, the tension with the Hubble constant direct measurement are at the $2.56\sigma$ level, $2.29\sigma$ level, $2.12\sigma$ level, $1.82\sigma$ level, $1.61\sigma$ level, and $1.60\sigma$ level, respectively.

From the constraint results, we find that compared with the $\Lambda$CDM model, the interacting dark energy plus massive (active/sterile) neutrino models can indeed relieve the tension between the Planck observation and the direct measurement from distance ladder (see also in Fig.~\ref{likeh01}). Particularly, when considering the dynamical dark energy in the interacting dark energy scenario, the tension can be relieved most significantly.
To show the effect of the dark energy properties on the constraint on $H_0$, the one-and two-dimensional posterior distributions contours of $H_0$ are shown in Figs.~\ref{vaiw}--\ref{vsiwh}.

We also find that, for the IDE models, the current LSS data can also affect the constraint results on $H_0$, and the values of $H_0$ would become lower when adding the LSS data in cosmological fit (see also Fig.~\ref{likeh01}).
Thus, for all the IDE models, the $H_0$ tension would be slightly larger with the addition of the LSS observational data.

In addition, in order to show how the massive (active/sterile) neutrinos affect the constraint results on $H_0$, we plot the one-dimensional posterior distributions of $H_0$ for the six considered models using CMB+BSH and CMB+BSH+LSS in Fig.~\ref{likeh02}.
From these figures, we can clearly see that for whichever data combination, the fitting values of $H_0$ in the IDE models with sterile neutrinos are always much larger than those with active neutrinos (see also Tables~\ref{tabactive}--\ref{tabsterile}). Thus, we can conclude that the IDE models with sterile neutrinos are more effective to relieve the $H_0$ tension.

In fact, although these extended models can alleviate the ``Hubble tension'' problem in some extent, it seems that they still cannot truly solve the problem because they are paying the price of using more parameters to fit the observational data (see also Refs. \cite{Zhao:2017urm,Guo:2018ans}).

\section{Conclusion}
\label{sec4}
In this paper, we have studied the interacting dynamical dark energy models with the energy transfer $Q=\beta H_0\rho_c$.
To discuss how the dynamical dark energy affects the massive (active/sterile) neutrino parameters in the IDE models, we take two typical interacting dynamical dark energy models as examples, i.e., the I$w$CDM model and the IHDE model, to make the analysis.
We use the PPF approach to calculate the perturbation of dark energy in the IDE cosmology.
The current observational data we use in this paper include the Planck 2015 CMB temperature and polarization data, the BAO data, the SN data, the $H_0$ direct measurement, the RSD data, the WL data, and the CMB lensing data.

For the IDE models with active neutrinos, we find that the dark energy properties could impact the constraint results on $\sum m_\nu$, and the IHDE+$\nu_a$ could give the most stringent upper limit on the total mass of active neutrinos among all the three models.
For the models with sterile neutrinos, we find that LSS data could significantly impact the constraints on the sterile neutrino parameters. What's more, the dark energy properties could also influence the constraint results on the sterile neutrino parameters.

By using the CMB+BSH data, we find that a negative $\beta$ is favored in the I$w$CDM model with massive (active/sterile) neutrinos and in this case $\beta=0$ is within the $1\sigma$ significance range, but the IHDE model with massive (active/sterile) neutrinos favor a positive $\beta$ ($\beta>0$) at more than $1\sigma$ significance level.
Thus, the dark energy properties could impact the constraint limits on the coupling parameter $\beta$.
By using the CMB+BSH+LSS data, we find that the current LSS data could also affect the constraints on $\beta$, and $\beta>0$ can be detected at $2.507\sigma$ significance for the IHDE+$\nu_s$ model.
Moreover, we also find that the consideration of massive (active/sterile) neutrinos in the interacting models could scarcely influence the constraint on the coupling parameter $\beta$.

In addition, we find that compared with the $\Lambda$CDM model, considering massive (active/sterile) neutrinos and interaction between dark sectors can indeed relieve the $H_0$ tension between the Planck observation and the Hubble constant direct measurement. In particular,
for the IHDE+$\nu_s$ model the $H_0$ tension can be reduced to be at the $1.28\sigma$ level by using the CMB+BSH data combination.
We also find that compared with the cases of active neutrinos, the consideration of sterile neutrinos in the IDE models could more effectively alleviate the $H_0$ tension.

\begin{acknowledgments}
This work was supported by the National Natural Science Foundation of China (Grant Nos. 11947022, 11975072, 11835009, 11875102, 11522540, and 11690021), the Liaoning Revitalization Talents Program (Grant No. XLYC1905011), the Fundamental Research Funds for the Central Universities (Grant No. N2005030), the National Program for Support of Top-Notch Young Professionals, the 2019 Annual Scientific Research Funding Project of the Education Department of Liaoning Province (Grant No. LJC201915), and the Doctoral Research Project of Shenyang Normal University (Grant No. BS201844).

\end{acknowledgments}


\begin{thebibliography}{99}
\bibitem{Lesgourgues:2006nd}
  J.~Lesgourgues and S.~Pastor,
  Phys.\ Rept.\  {\bf 429}, 307 (2006)
  [astro-ph/0603494].

\bibitem{Xing:2019vks}
  Z.~z.~Xing,
  arXiv:1909.09610 [hep-ph].

\bibitem{Abazajian:2012ys}
  K.~N.~Abazajian {\it et al.},
  arXiv:1204.5379 [hep-ph].
\bibitem{Hannestad:2012ky}
  S.~Hannestad, I.~Tamborra and T.~Tram,
  JCAP {\bf 1207}, 025 (2012)
  [arXiv:1204.5861 [astro-ph.CO]].
\bibitem{Conrad:2013mka}
  J.~M.~Conrad, W.~C.~Louis and M.~H.~Shaevitz,
  Ann.\ Rev.\ Nucl.\ Part.\ Sci.\  {\bf 63}, 45 (2013)
  [arXiv:1306.6494 [hep-ex]].

\bibitem{Aghanim:2015xee}
  N.~Aghanim {\it et al.} [Planck Collaboration],
  Astron.\ Astrophys.\  {\bf 594}, A11 (2016)
  [arXiv:1507.02704 [astro-ph.CO]].

\bibitem{Beutler:2011hx}
  F.~Beutler {\it et al.},
  Mon.\ Not.\ Roy.\ Astron.\ Soc.\  {\bf 416}, 3017 (2011)
  [arXiv:1106.3366 [astro-ph.CO]].
\bibitem{Ross:2014qpa}
  A.~J.~Ross, L.~Samushia, C.~Howlett, W.~J.~Percival, A.~Burden and M.~Manera,
  Mon.\ Not.\ Roy.\ Astron.\ Soc.\  {\bf 449}, no. 1, 835 (2015)
  [arXiv:1409.3242 [astro-ph.CO]].
\bibitem{Cuesta:2015mqa}
  A.~J.~Cuesta {\it et al.},
  Mon.\ Not.\ Roy.\ Astron.\ Soc.\  {\bf 457}, no. 2, 1770 (2016)
  [arXiv:1509.06371 [astro-ph.CO]].

\bibitem{Betoule:2014frx}
  M.~Betoule {\it et al.} [SDSS Collaboration],
  Astron.\ Astrophys.\  {\bf 568}, A22 (2014)
  [arXiv:1401.4064 [astro-ph.CO]].

\bibitem{Riess:2016jrr}
  A.~G.~Riess {\it et al.},
  Astrophys.\ J.\  {\bf 826}, no. 1, 56 (2016)
  [arXiv:1604.01424 [astro-ph.CO]].

\bibitem{Ade:2015zua}
  P.~A.~R.~Ade {\it et al.} [Planck Collaboration],
  Astron.\ Astrophys.\  {\bf 594}, A15 (2016)
  [arXiv:1502.01591 [astro-ph.CO]].

\bibitem{Zhang:2015uhk}
  X.~Zhang,
  Phys.\ Rev.\ D {\bf 93}, no. 8, 083011 (2016)
  [arXiv:1511.02651 [astro-ph.CO]].

\bibitem{Feng:2017mfs}
  L.~Feng, J.~F.~Zhang and X.~Zhang,
  Sci.\ China Phys.\ Mech.\ Astron.\  {\bf 61} (2018) no.5,  050411
  [arXiv:1706.06913 [astro-ph.CO]].


\bibitem{GonzalezGarcia:2007ib}
  M.~C.~Gonzalez-Garcia and M.~Maltoni,
  Phys.\ Rept.\  {\bf 460}, 1 (2008)
  [arXiv:0704.1800 [hep-ph]].
\bibitem{Li:2012vn}
  H.~Li and X.~Zhang,
  Phys.\ Lett.\ B {\bf 713}, 160 (2012)
  [arXiv:1202.4071 [astro-ph.CO]].
\bibitem{Wang:2012uf}
  Y.~H.~Li, S.~Wang, X.~D.~Li and X.~Zhang,
  JCAP {\bf 1302}, 033 (2013)
  [arXiv:1207.6679 [astro-ph.CO]].
\bibitem{Wang:2012vh}
  X.~Wang, X.~L.~Meng, T.~J.~Zhang, H.~Shan, Y.~Gong, C.~Tao, X.~Chen and Y.~F.~Huang,
  JCAP {\bf 1211}, 018 (2012)
  [arXiv:1210.2136 [astro-ph.CO]].
\bibitem{Jacques:2013xr}
  T.~D.~Jacques, L.~M.~Krauss and C.~Lunardini,
  Phys.\ Rev.\ D {\bf 87}, no. 8, 083515 (2013)
  [arXiv:1301.3119 [astro-ph.CO]].
\bibitem{Mirizzi:2013gnd}
  A.~Mirizzi, G.~Mangano, N.~Saviano, E.~Borriello, C.~Giunti, G.~Miele and O.~Pisanti,
  Phys.\ Lett.\ B {\bf 726}, 8 (2013)
  [arXiv:1303.5368 [astro-ph.CO]].
\bibitem{Drewes:2013gca}
  M.~Drewes,
  Int.\ J.\ Mod.\ Phys.\ E {\bf 22}, 1330019 (2013)
  [arXiv:1303.6912 [hep-ph]].
\bibitem{Hamann:2013iba}
  J.~Hamann and J.~Hasenkamp,
  JCAP {\bf 1310}, 044 (2013)
  [arXiv:1308.3255 [astro-ph.CO]].
\bibitem{Wyman:2013lza}
  M.~Wyman, D.~H.~Rudd, R.~A.~Vanderveld and W.~Hu,
  Phys.\ Rev.\ Lett.\  {\bf 112}, no. 5, 051302 (2014)
  [arXiv:1307.7715 [astro-ph.CO]].
\bibitem{Battye:2013xqa}
  R.~A.~Battye and A.~Moss,
  Phys.\ Rev.\ Lett.\  {\bf 112}, no. 5, 051303 (2014)
  [arXiv:1308.5870 [astro-ph.CO]].
\bibitem{Giusarma:2014zza}
  E.~Giusarma, E.~Di Valentino, M.~Lattanzi, A.~Melchiorri and O.~Mena,
  Phys.\ Rev.\ D {\bf 90}, no. 4, 043507 (2014)
  [arXiv:1403.4852 [astro-ph.CO]].
\bibitem{Dvorkin:2014lea}
  C.~Dvorkin, M.~Wyman, D.~H.~Rudd and W.~Hu,
  Phys.\ Rev.\ D {\bf 90}, no. 8, 083503 (2014)
  [arXiv:1403.8049 [astro-ph.CO]].
\bibitem{Lesgourgues:2014zoa}
  J.~Lesgourgues and S.~Pastor,
  New J.\ Phys.\  {\bf 16}, 065002 (2014)
  [arXiv:1404.1740 [hep-ph]].
\bibitem{Archidiacono:2014apa}
  M.~Archidiacono, N.~Fornengo, S.~Gariazzo, C.~Giunti, S.~Hannestad and M.~Laveder,
  JCAP {\bf 1406}, 031 (2014)
  [arXiv:1404.1794 [astro-ph.CO]].
\bibitem{Leistedt:2014sia}
  B.~Leistedt, H.~V.~Peiris and L.~Verde,
  Phys.\ Rev.\ Lett.\  {\bf 113}, 041301 (2014)
  [arXiv:1404.5950 [astro-ph.CO]].
\bibitem{Zhang:2014nta}
  J.~F.~Zhang, Y.~H.~Li and X.~Zhang,
  Eur.\ Phys.\ J.\ C {\bf 74}, 2954 (2014)
  [arXiv:1404.3598 [astro-ph.CO]].


\bibitem{Costanzi:2014tna}
  M.~Costanzi, B.~Sartoris, M.~Viel and S.~Borgani,
  JCAP {\bf 1410}, no. 10, 081 (2014)
  [arXiv:1407.8338 [astro-ph.CO]].
\bibitem{Zhang:2014lfa}
  J.~F.~Zhang, Y.~H.~Li and X.~Zhang,
  Phys.\ Lett.\ B {\bf 739}, 102 (2014)
  [arXiv:1408.4603 [astro-ph.CO]].
\bibitem{Zhang:2014dxk}
  J.~F.~Zhang, Y.~H.~Li and X.~Zhang,
  Phys.\ Lett.\ B {\bf 740}, 359 (2015)
  [arXiv:1403.7028 [astro-ph.CO]].
\bibitem{Zhu:2014qma}
  H.~M.~Zhu, U.~L.~Pen, X.~Chen and D.~Inman,
  Phys.\ Rev.\ Lett.\  {\bf 116}, no. 14, 141301 (2016)
  [arXiv:1412.1660 [astro-ph.CO]].
\bibitem{Li:2015poa}
  Y.~H.~Li, J.~F.~Zhang and X.~Zhang,
  Phys.\ Lett.\ B {\bf 744}, 213 (2015)
  [arXiv:1502.01136 [astro-ph.CO]].
\bibitem{Zhang:2015rha}
  J.~F.~Zhang, M.~M.~Zhao, Y.~H.~Li and X.~Zhang,
  JCAP {\bf 1504}, 038 (2015)
  [arXiv:1502.04028 [astro-ph.CO]].
\bibitem{Geng:2015haa}
  C.~Q.~Geng, C.~C.~Lee, R.~Myrzakulov, M.~Sami and E.~N.~Saridakis,
  JCAP {\bf 1601}, no. 01, 049 (2016)
  [arXiv:1504.08141 [astro-ph.CO]].
\bibitem{Qian:2015waa}
  X.~Qian and P.~Vogel,
  Prog.\ Part.\ Nucl.\ Phys.\  {\bf 83}, 1 (2015)
  [arXiv:1505.01891 [hep-ex]].
\bibitem{DellOro:2015kys}
  S.~Dell'Oro, S.~Marcocci, M.~Viel and F.~Vissani,
  JCAP {\bf 1512}, no. 12, 023 (2015)
  [arXiv:1505.02722 [hep-ph]].
\bibitem{Patterson:2015xja}
  R.~B.~Patterson,
  Ann.\ Rev.\ Nucl.\ Part.\ Sci.\  {\bf 65}, 177 (2015)
  [arXiv:1506.07917 [hep-ex]].
\bibitem{Gariazzo:2015rra}
  S.~Gariazzo, C.~Giunti, M.~Laveder, Y.~F.~Li and E.~M.~Zavanin,
  J.\ Phys.\ G {\bf 43}, 033001 (2016)
  [arXiv:1507.08204 [hep-ph]].
\bibitem{Huang:2015wrx}
  Q.~G.~Huang, K.~Wang and S.~Wang,
  Eur.\ Phys.\ J.\ C {\bf 76}, no. 9, 489 (2016)
  [arXiv:1512.05899 [astro-ph.CO]].
\bibitem{Capozzi:2016rtj}
  F.~Capozzi, E.~Lisi, A.~Marrone, D.~Montanino and A.~Palazzo,
  Nucl.\ Phys.\ B {\bf 908}, 218 (2016)
  [arXiv:1601.07777 [hep-ph]].
\bibitem{Giusarma:2016phn}
  E.~Giusarma, M.~Gerbino, O.~Mena, S.~Vagnozzi, S.~Ho and K.~Freese,
  Phys.\ Rev.\ D {\bf 94}, no. 8, 083522 (2016)
  [arXiv:1605.04320 [astro-ph.CO]].
\bibitem{DiValentino:2016hlg}
  E.~Di Valentino, A.~Melchiorri and J.~Silk,
  Phys.\ Lett.\ B {\bf 761}, 242 (2016)
  [arXiv:1606.00634 [astro-ph.CO]].
\bibitem{Lu:2016hsd}
  J.~Lu, M.~Liu, Y.~Wu, Y.~Wang and W.~Yang,
  Eur.\ Phys.\ J.\ C {\bf 76}, no. 12, 679 (2016)
  [arXiv:1606.02987 [astro-ph.CO]].
\bibitem{Wang:2016tsz}
  S.~Wang, Y.~F.~Wang, D.~M.~Xia and X.~Zhang,
  Phys.\ Rev.\ D {\bf 94}, no. 8, 083519 (2016)
  [arXiv:1608.00672 [astro-ph.CO]].
\bibitem{Zhao:2016ecj}
  M.~M.~Zhao, Y.~H.~Li, J.~F.~Zhang and X.~Zhang,
  Mon.\ Not.\ Roy.\ Astron.\ Soc.\  {\bf 469}, no. 2, 1713 (2017)
  [arXiv:1608.01219 [astro-ph.CO]].
\bibitem{Xu:2016ddc}
  L.~Xu and Q.~G.~Huang,
  Sci.\ China Phys.\ Mech.\ Astron.\  {\bf 61}, no. 3, 039521 (2018)
  [arXiv:1611.05178 [astro-ph.CO]].
\bibitem{Vagnozzi:2017ovm}
  S.~Vagnozzi, E.~Giusarma, O.~Mena, K.~Freese, M.~Gerbino, S.~Ho and M.~Lattanzi,
  Phys.\ Rev.\ D {\bf 96}, no. 12, 123503 (2017)
  [arXiv:1701.08172 [astro-ph.CO]].
\bibitem{Guo:2017hea}
  R.~Y.~Guo, Y.~H.~Li, J.~F.~Zhang and X.~Zhang,
  JCAP {\bf 1705}, no. 05, 040 (2017)
  [arXiv:1702.04189 [astro-ph.CO]].


\bibitem{Li:2017iur}
  E.~K.~Li, H.~Zhang, M.~Du, Z.~H.~Zhou and L.~Xu,
  JCAP {\bf 1808}, no. 08, 042 (2018)
  [arXiv:1703.01554 [astro-ph.CO]].
\bibitem{Yang:2017amu}
  W.~Yang, R.~C.~Nunes, S.~Pan and D.~F.~Mota,
  Phys.\ Rev.\ D {\bf 95}, no. 10, 103522 (2017)
  [arXiv:1703.02556 [astro-ph.CO]].
\bibitem{Capozzi:2017ipn}
  F.~Capozzi, E.~Di Valentino, E.~Lisi, A.~Marrone, A.~Melchiorri and A.~Palazzo,
  Phys.\ Rev.\ D {\bf 95}, no. 9, 096014 (2017)
  [arXiv:1703.04471 [hep-ph]].
\bibitem{Feng:2017nss}
  L.~Feng, J.~F.~Zhang and X.~Zhang,
  Eur.\ Phys.\ J.\ C {\bf 77}, no. 6, 418 (2017)
  [arXiv:1703.04884 [astro-ph.CO]].
\bibitem{Zhao:2017urm}
  M.~M.~Zhao, D.~Z.~He, J.~F.~Zhang and X.~Zhang,
  Phys.\ Rev.\ D {\bf 96}, no. 4, 043520 (2017)
  [arXiv:1703.08456 [astro-ph.CO]].
\bibitem{Abazajian:2017tcc}
  K.~N.~Abazajian,
  Phys.\ Rept.\  {\bf 711-712}, 1 (2017)
  [arXiv:1705.01837 [hep-ph]].
\bibitem{Wang:2017htc}
  S.~Wang, Y.~F.~Wang and D.~M.~Xia,
  Chin.\ Phys.\ C {\bf 42}, no. 6, 065103 (2018)
  [arXiv:1707.00588 [astro-ph.CO]].
\bibitem{Zhao:2017jma}
  M.~M.~Zhao, J.~F.~Zhang and X.~Zhang,
  Phys.\ Lett.\ B {\bf 779}, 473 (2018)
  [arXiv:1710.02391 [astro-ph.CO]].
\bibitem{Feng:2017usu}
  L.~Feng, J.~F.~Zhang and X.~Zhang,
  Phys.\ Dark Univ.\  {\bf 23}, 100261 (2019).
  [arXiv:1712.03148 [astro-ph.CO]].
\bibitem{Wang:2018lun}
  L.~F.~Wang, X.~N.~Zhang, J.~F.~Zhang and X.~Zhang,
  Phys.\ Lett.\ B {\bf 782}, 87 (2018)
  [arXiv:1802.04720 [astro-ph.CO]].
\bibitem{Guo:2018gyo}
  R.~Y.~Guo, J.~F.~Zhang and X.~Zhang,
  Chin.\ Phys.\ C {\bf 42}, no. 9, 095103 (2018)
  [arXiv:1803.06910 [astro-ph.CO]].
\bibitem{Guo:2018ans}
  R.~Y.~Guo, J.~F.~Zhang and X.~Zhang,
  JCAP {\bf 1902}, 054 (2019)
  [arXiv:1809.02340 [astro-ph.CO]].
\bibitem{Zhao:2018fjj}
  M.~M.~Zhao, R.~Y.~Guo, J.~F.~Zhang and X.~Zhang,
  Sci.\ China Phys.\ Mech.\ Astron.\  {\bf 63}, 230412 (2020),
  arXiv:1810.11658 [astro-ph.CO].
\bibitem{Feng:2019mym}
  L.~Feng, H.~L.~Li, J.~F.~Zhang and X.~Zhang,
  Sci.\ China Phys.\ Mech.\ Astron.\  {\bf 63}, 220401 (2020)
  [arXiv:1903.08848 [astro-ph.CO]].
\bibitem{Zhang:2019ipd}
  J.~F.~Zhang, B.~Wang and X.~Zhang,
  Sci.\ China Phys.\ Mech.\ Astron.\  {\bf 63}, 280411 (2020),
  arXiv:1907.00179 [astro-ph.CO].
\bibitem{Comelli:2003cv}
  D.~Comelli, M.~Pietroni and A.~Riotto,
  Phys.\ Lett.\ B {\bf 571}, 115 (2003)
  [hep-ph/0302080].
\bibitem{Cai:2004dk}
  R.~G.~Cai and A.~Wang,
  JCAP {\bf 0503}, 002 (2005)
  [hep-th/0411025].
\bibitem{Zhang:2004gc}
  X.~Zhang, F.~Q.~Wu and J.~Zhang,
  JCAP {\bf 0601}, 003 (2006)
  [astro-ph/0411221].
\bibitem{Zhang:2005rg}
  X.~Zhang,
  Mod.\ Phys.\ Lett.\ A {\bf 20}, 2575 (2005)
  [astro-ph/0503072].
\bibitem{Zimdahl:2005bk}
  W.~Zimdahl,
  Int.\ J.\ Mod.\ Phys.\ D {\bf 14}, 2319 (2005)
  [gr-qc/0505056].
\bibitem{Wang:2006qw}
  B.~Wang, J.~Zang, C.~Y.~Lin, E.~Abdalla and S.~Micheletti,
  Nucl.\ Phys.\ B {\bf 778}, 69 (2007)
  [astro-ph/0607126].
\bibitem{Guo:2007zk}
  Z.~K.~Guo, N.~Ohta and S.~Tsujikawa,
  Phys.\ Rev.\ D {\bf 76}, 023508 (2007)
  [astro-ph/0702015 [ASTRO-PH]].
\bibitem{Zhang:2007uh}
  J.~Zhang, H.~Liu and X.~Zhang,
  Phys.\ Lett.\ B {\bf 659}, 26 (2008)
  [arXiv:0705.4145 [astro-ph]].
\bibitem{He:2008tn}
  J.~H.~He and B.~Wang,
  JCAP {\bf 0806}, 010 (2008)
  [arXiv:0801.4233 [astro-ph]].
\bibitem{Li:2009zs}
  M.~Li, X.~D.~Li, S.~Wang, Y.~Wang and X.~Zhang,
  JCAP {\bf 0912}, 014 (2009)
  [arXiv:0910.3855 [astro-ph.CO]].
\bibitem{Zhang:2009qa}
  L.~Zhang, J.~Cui, J.~Zhang and X.~Zhang,
  Int.\ J.\ Mod.\ Phys.\ D {\bf 19}, 21 (2010)
  [arXiv:0911.2838 [astro-ph.CO]].


\bibitem{He:2010im}
  J.~H.~He, B.~Wang and E.~Abdalla,
  Phys.\ Rev.\ D {\bf 83}, 063515 (2011)
  [arXiv:1012.3904 [astro-ph.CO]].
\bibitem{Chen:2011rz}
  Y.~Chen, Z.~H.~Zhu, L.~Xu and J.~S.~Alcaniz,
  Phys.\ Lett.\ B {\bf 698}, 175 (2011)
  [arXiv:1103.2512 [astro-ph.CO]].
\bibitem{Li:2011ga}
  Y.~H.~Li and X.~Zhang,
  Eur.\ Phys.\ J.\ C {\bf 71}, 1700 (2011)
  [arXiv:1103.3185 [astro-ph.CO]].
\bibitem{Clemson:2011an}
  T.~Clemson, K.~Koyama, G.~B.~Zhao, R.~Maartens and J.~Valiviita,
  Phys.\ Rev.\ D {\bf 85}, 043007 (2012)
  [arXiv:1109.6234 [astro-ph.CO]].
\bibitem{Fu:2011ab}
  T.~F.~Fu, J.~F.~Zhang, J.~Q.~Chen and X.~Zhang,
  Eur.\ Phys.\ J.\ C {\bf 72}, 1932 (2012)
  [arXiv:1112.2350 [astro-ph.CO]].
\bibitem{Zhang:2012uu}
  Z.~Zhang, S.~Li, X.~D.~Li, X.~Zhang and M.~Li,
  JCAP {\bf 1206}, 009 (2012)
  [arXiv:1204.6135 [astro-ph.CO]].
\bibitem{Salvatelli:2013wra}
  V.~Salvatelli, A.~Marchini, L.~Lopez-Honorez and O.~Mena,
  Phys.\ Rev.\ D {\bf 88}, no. 2, 023531 (2013)
  [arXiv:1304.7119 [astro-ph.CO]].


\bibitem{Xu:2013jma}
  X.~D.~Xu, B.~Wang, P.~Zhang and F.~Atrio-Barandela,
  JCAP {\bf 1312}, 001 (2013)
  [arXiv:1308.1475 [astro-ph.CO]].
\bibitem{Zhang:2013zyn}
  M.~J.~Zhang and W.~B.~Liu,
  Eur.\ Phys.\ J.\ C {\bf 74}, 2863 (2014)
  [arXiv:1312.0224 [astro-ph.CO]].
\bibitem{Li:2013bya}
  Y.~H.~Li and X.~Zhang,
  Phys.\ Rev.\ D {\bf 89}, no. 8, 083009 (2014)
  [arXiv:1312.6328 [astro-ph.CO]].
\bibitem{Yang:2014gza}
  W.~Yang and L.~Xu,
  Phys.\ Rev.\ D {\bf 89}, no. 8, 083517 (2014)
  [arXiv:1401.1286 [astro-ph.CO]].
\bibitem{yang:2014vza}
  W.~Yang and L.~Xu,
  JCAP {\bf 1408}, 034 (2014)
  [arXiv:1401.5177 [astro-ph.CO]].
\bibitem{Geng:2015ara}
  J.~J.~Geng, Y.~H.~Li, J.~F.~Zhang and X.~Zhang,
  Eur.\ Phys.\ J.\ C {\bf 75}, no. 8, 356 (2015)
  [arXiv:1501.03874 [astro-ph.CO]].
\bibitem{Duniya:2015nva}
  D.~G.~A.~Duniya, D.~Bertacca and R.~Maartens,
  Phys.\ Rev.\ D {\bf 91}, 063530 (2015)
  [arXiv:1502.06424 [astro-ph.CO]].
\bibitem{Yin:2015pqa}
  J.~L.~Cui, L.~Yin, L.~F.~Wang, Y.~H.~Li and X.~Zhang,
  JCAP {\bf 1509}, no. 09, 024 (2015)
  [arXiv:1503.08948 [astro-ph.CO]].
\bibitem{Murgia:2016ccp}
  R.~Murgia, S.~Gariazzo and N.~Fornengo,
  JCAP {\bf 1604}, no. 04, 014 (2016)
  [arXiv:1602.01765 [astro-ph.CO]].
\bibitem{Wang:2016lxa}
  B.~Wang, E.~Abdalla, F.~Atrio-Barandela and D.~Pavon,
  Rept.\ Prog.\ Phys.\  {\bf 79}, no. 9, 096901 (2016)
  [arXiv:1603.08299 [astro-ph.CO]].
\bibitem{Pourtsidou:2016ico}
  A.~Pourtsidou and T.~Tram,
  Phys.\ Rev.\ D {\bf 94}, no. 4, 043518 (2016)
  [arXiv:1604.04222 [astro-ph.CO]].
\bibitem{Costa:2016tpb}
  A.~A.~Costa, X.~D.~Xu, B.~Wang and E.~Abdalla,
  JCAP {\bf 1701}, no. 01, 028 (2017)
  [arXiv:1605.04138 [astro-ph.CO]].
\bibitem{Feng:2016djj}
  L.~Feng and X.~Zhang,
  JCAP {\bf 1608}, no. 08, 072 (2016)
  [arXiv:1607.05567 [astro-ph.CO]].
\bibitem{Xia:2016vnp}
  D.~M.~Xia and S.~Wang,
  Mon.\ Not.\ Roy.\ Astron.\ Soc.\  {\bf 463}, no. 1, 952 (2016)
  [arXiv:1608.04545 [astro-ph.CO]].
\bibitem{vandeBruck:2016hpz}
  C.~van de Bruck, J.~Mifsud and J.~Morrice,
  Phys.\ Rev.\ D {\bf 95}, no. 4, 043513 (2017)
  [arXiv:1609.09855 [astro-ph.CO]].
\bibitem{Sola:2017jbl}
   J.~Sol\`{a} Peracaula, J.~d.~C.~Perez and A.~Gomez-Valent,
  Mon.\ Not.\ Roy.\ Astron.\ Soc.\  {\bf 478}, no. 4, 4357 (2018)
  [arXiv:1703.08218 [astro-ph.CO]].
\bibitem{Yang:2018euj}
  W.~Yang, S.~Pan, E.~Di Valentino, R.~C.~Nunes, S.~Vagnozzi and D.~F.~Mota,
  JCAP {\bf 1809}, no. 09, 019 (2018)
  [arXiv:1805.08252 [astro-ph.CO]].
\bibitem{Guo:2017deu}
  J.~J.~Guo, J.~F.~Zhang, Y.~H.~Li, D.~Z.~He and X.~Zhang,
  Sci.\ China Phys.\ Mech.\ Astron.\  {\bf 61}, no. 3, 030011 (2018)
  [arXiv:1710.03068 [astro-ph.CO]].
\bibitem{Li:2018ydj}
  H.~L.~Li, L.~Feng, J.~F.~Zhang and X.~Zhang,
  Sci.\ China Phys.\ Mech.\ Astron.\  {\bf 62}, no. 12, 120411 (2019)
  [arXiv:1812.00319 [astro-ph.CO]].


\bibitem{Zhang:2005hs}
  X.~Zhang and F.~Q.~Wu,
  Phys.\ Rev.\ D {\bf 72}, 043524 (2005).
 [astro-ph/0506310].

\bibitem{Chang:2005ph}
  Z.~Chang, F.~Q.~Wu and X.~Zhang,
  Phys.\ Lett.\ B {\bf 633}, 14 (2006).
  [astro-ph/0509531].

\bibitem{Zhang:2006qu}
  X.~Zhang,
  Phys.\ Rev.\ D {\bf 74}, 103505 (2006).
  [astro-ph/0609699].

\bibitem{Zhang:2006av}
  X.~Zhang,
  Phys.\ Lett.\ B {\bf 648}, 1 (2007).
  [astro-ph/0604484].

\bibitem{Zhang:2007sh}
  X.~Zhang and F.~Q.~Wu,
  Phys.\ Rev.\ D {\bf 76}, 023502 (2007).
  [astro-ph/0701405].

\bibitem{Zhang:2007es}
  J.~Zhang, X.~Zhang and H.~Liu,
  Phys.\ Lett.\ B {\bf 651}, 84 (2007).
  [arXiv:0706.1185 [astro-ph]].

\bibitem{Zhang:2007an}
  J.~F.~Zhang, X.~Zhang and H.~Y.~Liu,
  Eur.\ Phys.\ J.\ C {\bf 52}, 693 (2007).
  [arXiv:0708.3121 [hep-th]].

\bibitem{Ma:2007av}
  Y.~Z.~Ma and X.~Zhang,
  Phys.\ Lett.\ B {\bf 661}, 239 (2008).
  [arXiv:0709.1517 [astro-ph]].

\bibitem{Li:2009bn}
  M.~Li, X.~D.~Li, S.~Wang and X.~Zhang,
  JCAP {\bf 0906}, 036 (2009).
  [arXiv:0904.0928 [astro-ph.CO]].

\bibitem{Zhang:2009xj}
  X.~Zhang,
  Phys.\ Lett.\ B {\bf 683}, 81 (2010).
  [arXiv:0909.4940 [gr-qc]].

\bibitem{Cui:2010dr}
  J.~Cui and X.~Zhang,
  Phys.\ Lett.\ B {\bf 690}, 233 (2010).
  [arXiv:1005.3587 [astro-ph.CO]].

\bibitem{Bamba:2012cp}
  K.~Bamba, S.~Capozziello, S.~Nojiri and S.~D.~Odintsov,
  Astrophys.\ Space Sci.\  {\bf 342}, 155 (2012).
 [arXiv:1205.3421 [gr-qc]].

\bibitem{Wang:2016och}
  S.~Wang, Y.~Wang and M.~Li,
  Phys.\ Rept.\  {\bf 696}, 1 (2017).
  [arXiv:1612.00345 [astro-ph.CO]].


\bibitem{Zhao:2005vj}
  G.~B.~Zhao, J.~Q.~Xia, M.~Li, B.~Feng and X.~Zhang,
  Phys.\ Rev.\ D {\bf 72}, 123515 (2005).
  [astro-ph/0507482].
\bibitem{Valiviita:2008iv}
  J.~Valiviita, E.~Majerotto and R.~Maartens,
  JCAP {\bf 0807}, 020 (2008).
  [arXiv:0804.0232 [astro-ph]].
\bibitem{He:2008si}
  J.~H.~He, B.~Wang and E.~Abdalla,
  Phys.\ Lett.\ B {\bf 671}, 139 (2009)
  [arXiv:0807.3471 [gr-qc]].

\bibitem{Li:2014eha}
  Y.~H.~Li, J.~F.~Zhang and X.~Zhang,
  Phys.\ Rev.\ D {\bf 90}, no. 6, 063005 (2014)
  [arXiv:1404.5220 [astro-ph.CO]].
\bibitem{Li:2014cee}
  Y.~H.~Li, J.~F.~Zhang and X.~Zhang,
  Phys.\ Rev.\ D {\bf 90}, no. 12, 123007 (2014)
  [arXiv:1409.7205 [astro-ph.CO]].
\bibitem{Li:2015vla}
  Y.~H.~Li, J.~F.~Zhang and X.~Zhang,
  Phys.\ Rev.\ D {\bf 93}, no. 2, 023002 (2016)
  [arXiv:1506.06349 [astro-ph.CO]].


\bibitem{Hu:2008zd}
  W.~Hu,
  Phys.\ Rev.\ D {\bf 77}, 103524 (2008).
  [arXiv:0801.2433 [astro-ph]].
\bibitem{Fang:2008sn}
  W.~Fang, W.~Hu and A.~Lewis,
  Phys.\ Rev.\ D {\bf 78}, 087303 (2008).
 [arXiv:0808.3125 [astro-ph]].

\bibitem{Lewis:2002ah}
  A.~Lewis and S.~Bridle,
  Phys.\ Rev.\ D {\bf 66}, 103511 (2002).
  [astro-ph/0205436].

\bibitem{Gil-Marin:2016wya}
   H.~Gil-Mar\'{\i}n, W.~J.~Percival, L.~Verde, J.~R.~Brownstein, C.~H.~Chuang, F.~S.~Kitaura, S.~A.~Rodr\'{\i}guez-Torres and M.~D.~Olmstead,
  Mon.\ Not.\ Roy.\ Astron.\ Soc.\  {\bf 465}, no. 2, 1757 (2017)
  [arXiv:1606.00439 [astro-ph.CO]].

\bibitem{Heymans:2013fya}
  C.~Heymans {\it et al.},
  Mon.\ Not.\ Roy.\ Astron.\ Soc.\  {\bf 432}, 2433 (2013)
  [arXiv:1303.1808 [astro-ph.CO]].



\end{thebibliography}
\end{document}